\documentclass{article}

\def\hybrid{\topmargin -20pt    \oddsidemargin 0pt
        \headheight 0pt \headsep 0pt
        \textwidth 6.25in       % A4 paper
        \textheight 9.5in       % A4 paper
        \marginparwidth .875in
        \parskip 5pt plus 1pt   \jot = 1.5ex}

\hybrid

\catcode`\@=11

\def\marginnote#1{}
\newcount\hour
\newcount\minute
\newtoks\amorpm
\hour=\time\divide\hour by60
\minute=\time{\multiply\hour by60 \global\advance\minute by-\hour}
\edef\standardtime{{\ifnum\hour<12 \global\amorpm={am}%
        \else\global\amorpm={pm}\advance\hour by-12 \fi
        \ifnum\hour=0 \hour=12 \fi
        \number\hour:\ifnum\minute<10 0\fi\number\minute\the\amorpm}}
\edef\militarytime{\number\hour:\ifnum\minute<10 0\fi\number\minute}
\def\draftlabel#1{{\@bsphack\if@filesw {\let\thepage\relax
   \xdef\@gtempa{\write\@auxout{\string
      \newlabel{#1}{{\@currentlabel}{\thepage}}}}}\@gtempa
   \if@nobreak \ifvmode\nobreak\fi\fi\fi\@esphack}
        \gdef\@eqnlabel{#1}}
\def\@eqnlabel{}
\def\@vacuum{}
\def\draftmarginnote#1{\marginpar{\raggedright\scriptsize\tt#1}}

\def\draft{\oddsidemargin -.5truein
        \def\@oddfoot{\sl preliminary draft \hfil
        \rm\thepage\hfil\sl\today\quad\militarytime}
        \let\@evenfoot\@oddfoot \overfullrule 3pt
        \let\label=\draftlabel
        \let\marginnote=\draftmarginnote
   \def\@eqnnum{(\theequation)\rlap{\kern\marginparsep\tt\@eqnlabel}%
\global\let\@eqnlabel\@vacuum}  }

\def\preprint{\twocolumn\sloppy\flushbottom\parindent 2em
        \leftmargini 2em\leftmarginv .5em\leftmarginvi .5em
        \oddsidemargin -.5in    \evensidemargin -.5in
        \columnsep .4in \footheight 0pt
        \textwidth 10.in        \topmargin  -.4in
        \headheight 12pt \topskip .4in
        \textheight 6.9in \footskip 0pt
        \def\@oddhead{\thepage\hfil\addtocounter{page}{1}\thepage}
        \let\@evenhead\@oddhead \def\@oddfoot{} \def\@evenfoot{} }

\def\numberbysection{\@addtoreset{equation}{section}
        \def\theequation{\thesection.\arabic{equation}}}

\def\underline#1{\relax\ifmmode\@@underline#1\else
        $\@@underline{\hbox{#1}}$\relax\fi}

\def\titlepage{\@restonecolfalse\if@twocolumn\@restonecoltrue\onecolumn
     \else \newpage \fi \thispagestyle{empty}\c@page\z@
        \def\thefootnote{\fnsymbol{footnote}} }

\def\endtitlepage{\if@restonecol\twocolumn \else \newpage \fi
        \def\thefootnote{\arabic{footnote}}
        \setcounter{footnote}{0}}  %\c@footnote\z@ }

\catcode`@=12
\relax

\makeatletter
\newcounter{pubctr}
\def\publist{\@ifnextchar[{\@publist}{\@@publist}}
\def\@publist[#1]{\list
        {[\arabic{pubctr}]\hfill}{\settowidth\labelwidth{[999]}
        \leftmargin\labelwidth
        \advance\leftmargin\labelsep
        \@nmbrlisttrue\def\@listctr{pubctr}
        \setcounter{pubctr}{#1}\addtocounter{pubctr}{-1}}}
\def\@@publist{\list
        {[\arabic{pubctr}]\hfill}{\settowidth\labelwidth{[999]}
        \leftmargin\labelwidth
        \advance\leftmargin\labelsep
        \@nmbrlisttrue\def\@listctr{pubctr}}}
 \relax
\makeatother
\newskip\humongous \humongous=0pt plus 1000pt minus 1000pt

\newif\ifdtup

\relax

\def\d{\partial}

\def\sqr#1#2{{\vcenter{\vbox{\hrule height.#2pt\hbox{\vrule width.#2pt 
height#1pt \kern#1pt \vrule width.#2pt}\hrule height.#2pt}}}}

\def\=d{\,{\buildrel\rm def\over =}\,}

\def\i3p{\p32\int d^3p}

\def\As{A\hbox to 1pt{\hss /}}
\def\np4{\int d^4p_1\cdots d^4p_{n-1}\, }

\def\nx4{\int d^4x_1\ldots d^4x_n\, }

\def\kon#1#2{\vbox{\halign{##&&##\cr
\lower4pt\hbox{$\scriptscriptstyle\vert$}\hrulefill &
\hrulefill\lower4pt\hbox{$\scriptscriptstyle\vert$}\cr $#1$&
$#2$\cr}}}

\def\konv#1#2#3{\hbox{\vrule height12pt depth-1pt}
\vbox{\hrule height12pt width#1cm depth-11.6pt}
\hbox{\vrule height6.5pt depth-0.5pt}
\vbox{\hrule height11pt width#2cm depth-10.6pt\kern5pt
      \hrule height6.5pt width#2cm depth-6.1pt}
\hbox{\vrule height12pt depth-1pt}
\vbox{\hrule height6.5pt width#3cm depth-6.1pt}
\hbox{\vrule height6.5pt depth-0.5pt}}
\def\konu#1#2#3{\hbox{\vrule height12pt depth-1pt}
\vbox{\hrule height1pt width#1cm depth-0.6pt}
\hbox{\vrule height12pt depth-6.5pt}
\vbox{\hrule height6pt width#2cm depth-5.6pt\kern5pt
      \hrule height1pt width#2cm depth-0.6pt}
\hbox{\vrule height12pt depth-6.5pt}
\vbox{\hrule height1pt width#3cm depth-0.6pt}
\hbox{\vrule height12pt depth-1pt}}

\def\konw#1#2#3{\hbox{\vrule height12pt depth-1pt}
\vbox{\hrule height12pt width#1cm depth-11.6pt}
\hbox{\vrule height6.5pt depth-0.5pt}
\vbox{\hrule height12pt width#2cm depth-11.6pt \kern5pt
      \hrule height6.5pt width#2cm depth-6.1pt}
\hbox{\vrule height6.5pt depth-0.5pt}
\vbox{\hrule height12pt width#3cm depth-11.6pt}
\hbox{\vrule height12pt depth-1pt}}

\def\i{{\rm int}}

\def\c{{\rm cl}}
\def\e{{\rm ext}}

\def\r{{\rm ret}}
\def\a{{\rm av}}

\def\m3{{\mu_1\mu_2\mu_3}}

\def\p{{(+)}}

\def\be{\begin{equation}}       \def\eq{\begin{equation}}
\def\ee{\end{equation}}         \def\eqe{\end{equation}}

\def\bea{\begin{eqnarray}}      \def\eqa{\begin{eqnarray}}
\def\ena{\end{eqnarray}}        \def\eea{\end{eqnarray}}
                                \def\eqae{\end{eqnarray}}

\def\ba{\begin{array}}
\def\ea{\end{array}}
\def\unit{1 \hskip-.3em \raise2pt\hbox{$ \scriptstyle |$ } }
\def\a{\alpha}
\def\b{\beta}
\def\c{\gamma} 
\def\d{\delta}
\def\e{\epsilon}           % Also, \varepsilon
\def\f{\phi}               %      \varphi
\def\g{\gamma}
   
\def\i{\iota}
\def\j{\psi}
\def\l{\lambda}
\def\m{\mu}
\def\n{\nu}
  
\def\p{\pi}                % Also, \varpi
\def\r{\rho}                                     %     \varrho
\def\s{\sigma}                                   %     \varsigma
\def\t{\tau}

\def\G{\Gamma}

\def\L{\Lambda}
\def\O{\Omega}

\def\ca{{\cal A}}

\def\cn{{\cal N}}

\def\cq{{\cal Q}}

\def\half{{1 \over 2}}

\def\bop#1{\setbox0=\hbox{$#1M$}\mkern1.5mu
        \vbox{\hrule height0pt depth.04\ht0
        \hbox{\vrule width.04\ht0 height.9\ht0 \kern.9\ht0
        \vrule width.04\ht0}\hrule height.04\ht0}\mkern1.5mu}
\def\>{\rangle} %right angle

\def\<{\langle} %left angle
\def\Dsl{D \hskip-.6em \raise1pt\hbox{$ / $ } }

\def\sl#1{\rlap{\hbox{$\mskip 1 mu /$}}#1}% good slash for l.c.
\def\leftrightarrowfill{$\mathsurround=0pt \mathord\leftarrow \mkern-6mu
       \cleaders\hbox{$\mkern-2mu \mathord- \mkern-2mu$}\hfill
       \mkern-6mu \mathord\rightarrow$}
\def\dvec#1{\vbox{\ialign{##\crcr
       \leftrightarrowfill\crcr\noalign{\kern-1pt\nointerlineskip}
       $\hfil\displaystyle{#1}\hfil$\crcr}}}          % <--> accent
\def\hook#1{{\vrule height#1pt width0.4pt depth0pt}}
\def\leftrighthookfill#1{$\mathsurround=0pt \mathord\hook#1
       \hrulefill\mathord\hook#1$}
\def\underhook#1{\vtop{\ialign{##\crcr                 % |_| under
       $\hfil\displaystyle{#1}\hfil$\crcr
       \noalign{\kern-1pt\nointerlineskip\vskip2pt}
       \leftrighthookfill5\crcr}}}
\def\smallunderhook#1{\vtop{\ialign{##\crcr      % " for su'scripts
       $\hfil\scriptstyle{#1}\hfil$\crcr
       \noalign{\kern-1pt\nointerlineskip\vskip2pt}
       \leftrighthookfill3\crcr}}}
\def\sfrac#1#2{{\vphantom1\smash{\lower.5ex\hbox{\small$#1$}}\over
       \vphantom1\smash{\raise.4ex\hbox{\small$#2$}}}} % alt. fraction
\def\bfrac#1#2{{\vphantom1\smash{\lower.5ex\hbox{$#1$}}\over
       \vphantom1\smash{\raise.3ex\hbox{$#2$}}}}      % "
\def\afrac#1#2{{\vphantom1\smash{\lower.5ex\hbox{$#1$}}\over#2}}  %"
\def\on#1#2{{\buildrel{\mkern2.5mu#1\mkern-2.5mu}\over{#2}}}%acc.over
\def\ddt#1{\on{\hbox{\LARGE .\kern-2pt.}}#1}             % double dot
\def\tdt#1{\on{\hbox{\LARGE .\kern-2pt.\kern-2pt.}}#1}   % triple dot

\def\boxes#1{
       \newcount\num
       \num=1
       \newdimen\downsy
       \downsy=-1.5ex
       \mskip-2.8mu
       \bo
       \loop
       \ifnum\num<#1
       \llap{\raise\num\downsy\hbox{$\bo$}}
       \advance\num by1
       \repeat}
\def\boxup#1#2{\newcount\numup
       \numup=#1
       \advance\numup by-1
       \newdimen\upsy
       \upsy=.75ex
       \mskip2.8mu
       \raise\numup\upsy\hbox{$#2$}}

\newskip\humongous \humongous=0pt plus 1000pt minus 1000pt

\newif\ifdtup

\def\PRD{Phys. Rev. D}
\def\PRL#1#2#3{{\it Phys. Rev. Lett.} {\bf#1} (#2) #3}

\def\NPB#1#2#3{{\it Nucl. Phys.} {\bf B#1} (#2) #3}

\def\PRD#1#2#3{{\it Phys. Rev.} {\bf D#1} (#2) #3}

\def\PLB#1#2#3{{\it Phys. Lett.} {\bf #1B} (#2) #3}

\def\PRep#1#2#3{{\it Phys. Reports} {\bf #1} (#2) #3}

\def\to{\rightarrow}

\def\1ov4{{1\over 4}}

\def\xx{\times}

\def\ddt{\dot{\t}}

\def\xx{\times}

\def\nonu{\nonumber \\{}}
\def\half{{1 \over 2}}

\def\ot{\otimes}
\def\id{{\bf 1}}

\begin{document}

\thispagestyle{empty}
\begin{flushright}
{\sc KUL-TF}-98/17\\
hep-th/9803231
\end{flushright}
\vspace{3cm}
\setcounter{footnote}{0}
\begin{center}
{\LARGE\sc{Brane intersections, anti-de Sitter spacetimes and
dual superconformal theories
    }}\\[14mm]

\sc{Harm Jan Boonstra\footnote{e-mail: harm.boonstra@fys.kuleuven.ac.be},
    Bas Peeters\footnote{e-mail: bas.peeters@fys.kuleuven.ac.be} and
    Kostas Skenderis\footnote{e-mail:
kostas.skenderis@fys.kuleuven.ac.be},}\\[5mm] 
{\it Instituut voor Theoretische Fysica\\
KU Leuven\\
Celestijnenlaan 200D\\
3001 Heverlee, Belgium}\\[20mm]

{\sc Abstract}\\[2mm]

\end{center}

We construct a class of intersecting brane solutions
with horizon geometries of the form $adS_k \xx S^l \xx S^m \xx E^n$.
We describe how all these solutions are connected
through the addition of a wave and/or monopoles.
All solutions exhibit supersymmetry enhancement near the horizon.
Furthermore we argue
that string theory on these spaces is dual to 
specific superconformal field theories in two dimensions whose
symmetry algebra in all cases contains the large $\cn=4$ algebra $\ca_\g$.
Implications for gauged supergravities are also discussed.

\vfill

\newpage

\section{Introduction}

During the last few years a new unifying picture of all string theories
has emerged. All of them can be viewed as special limits
of an eleven-dimensional theory, M-theory\cite{mtheory}.
M-theory at low energies is described by eleven-dimensional 
supergravity ($11d$ SUGRA)\cite{11sugra}. This has led to an extensive  
study of the latter. However, most of the studies were confined to 
toroidal compactifications of $11d$ SUGRA.  
One reason for this is that in this case the connection with 
string theories is rather direct. These compactifications 
lead to Poincar\'{e} supergravities in lower dimensions. From 
the point of view of M-theory, however, there is no {\it a priori} 
reason that distinguishes toroidal compactifications from other ones.
A large class of compactifications that received considerable attention
in the mid-eighties are compactifications on spheres. These compactifications
lead to (maximally supersymmetric) gauged supergravities in lower dimensions
\cite{S7gsugra,gaugedsu,DNP}. 
The latter contain a negative cosmological constant in the action and in 
many cases admit a stable anti-de Sitter (adS) vacuum. A famous example 
is the compactification of eleven-dimensional supergravity on $S^7$, 
which yields \cite{S7gsugra} $\cn=8$ $adS_4$ gauged 
supergravity\cite{N8gaugesugra}. 

A natural question is what is the r\^{o}le of gauged supergravities 
in M-theory. Since the asymptotic group associated to these theories
is different from the one associated to Poincar\'{e} supergravities 
one may consider them as different sectors of 
M-theory. However, one of the lessons that our experience with dualities 
teaches us is that differently looking theories may actually be 
equivalent. Indeed, it has been known for some time that dualities can 
change the asymptotic geometry of spacetime\cite{topch}.
We have recently\cite{BPS1} provided a connection between Poincar\'{e} and 
gauged supergravities by giving a set of duality transformations
(such duality transformations have also appeared in \cite{hyun}
and were recently also studied in \cite{berg})  
that map solutions of the former to solutions of the latter.
In particular, we have shown that certain brane configurations
and intersections thereof which are asymptotically flat are 
mapped by the so-called shift transformation\cite{BPS1} to solutions
that are locally isometric to $adS_k \xx E^l \xx S^m$,
where $adS_k$ is the $k$-dimensional anti-de Sitter space, $E^l$ is
the $l$-dimensional Euclidean space and $S^m$ is the
$m$-dimensional sphere. Based on these considerations, 
it was further suggested in \cite{KSKS} that the actual symmetry group 
of M-theory is bigger than what one usually assumes, and allows
for connections between spacetimes with different asymptotic
group and different number of (non-compact) dimensions. 
The fact that these considerations lead to a microscopic derivation 
of the Bekenstein-Hawking entropy formula for $4d$ and $5d$ non-extremal
black holes and organize many results on black hole 
entropy in a unifying way\cite{KSKS} strongly supports this point of view.

The proposed new dualities are rather surprising from the spacetime 
point of view. In a given spacetime, the particle states carry unitary 
irreducible representations (UIRs) of the spacetime group. If the spacetime
is asymptotically flat then the particle states carry UIRs of the 
Poincar\'{e} group. After the duality transformation the spacetime 
changes to one which is asymptotically anti-de Sitter. So, now
the particle states should carry UIRs of the anti-de Sitter group.
Therefore, a necessary condition for the duality to work is that 
there are multiplets of the anti-de Sitter
group that can carry the original degrees of freedom. This is a 
very non-trivial condition since from the point of view of classical 
relativity asymptotically flat spacetimes are different from 
asymptotically anti-de Sitter spacetimes. It is string theory
that makes such a connection possible.

To make the discussion concrete consider the case of 
a Dirichlet three-brane ($D3$) in
IIB string theory. The duality transformation maps the supergravity 
solution describing the $D3$ brane to a space 
which is locally isometric to $adS_5 \xx S^5$. In the limit of decoupling
gravity one is left with the $D3$ worldvolume fields. For a single
$D3$ brane the latter belong to a $U(1)$ $\cn=4$ super Yang-Mills (SYM) 
multiplet. For the duality to work there should exist a UIR of
the $adS_5$ supergroup that contains precisely these fields.
Indeed, it turns out that such a multiplet exists. It is the 
so-called doubleton multiplet of $adS_5$.\footnote{
Actually the doubleton multiplet consists of six real scalars, 
four complex spinors and a complex antisymmetric tensor\cite{doub1, doub2}. 
A tensor in five dimensions is dual to a vector. Since the tensor 
is pure gauge in the bulk, so is the dual vector field. So, at the 
end one obtains an $\cn=4$ $U(1)$ SYM multiplet on the boundary.
As far as we know, it is not known how to combine antisymmetric
tensor gauge invariance with Yang-Mills gauge invariance. Thus, it is not 
clear whether one can associate $\cn=4$ $SU(N)$ SYM theory with 
singletons. \label{nonab}} 
The latter is the 
most fundamental multiplet of $adS_5$. All other UIRs can be constructed 
by a tensoring procedure\cite{doub1}. The doubleton multiplet appeared in the 
compactification of IIB supergravity on $S^5$\cite{doub2}. However,
these degrees of freedom could be gauged away everywhere but in the 
boundary of $adS_5$. The anti-de Sitter group $SO(d-1,2)$ coincides 
with the conformal group in one dimension lower. It follows that the 
singleton field theory is a superconformal field theory. 
Similar remarks apply for the M-theory branes. The supergravity solution
describing a membrane $M2$ (a fivebrane $M5$) is mapped to a spacetime which 
is locally isometric to $adS_4 \xx S^7$ ($adS_7 \xx S^4$), and furthermore the
worldvolume fields of $M2$ ($M5$) are the same as the ones in the singleton 
(doubleton) multiplet of $adS_4$ ($adS_7$). (From now on, we shall not 
distinguish between singletons and doubletons. We shall call both 
singletons.\footnote{Singletons were discovered by Dirac
(for $adS_4$), and named singletons by Fronsdal \cite{FF}.})
These facts have led to an association of branes to 
singletons\cite{m2,NST,m5,f1,KSKS,BPS2}.

A puzzle arises, however, when one considers multiple coincident 
branes. In the case of $D$-branes there are new massless
states that arise from strings that become massless when one moves
the branes on top of each other. This leads to enhanced gauge symmetry
\cite{witten1}. In particular, the worldvolume theory of $N$ coincident
$D3$ branes is an $SU(N)$ $\cn=4$ SYM theory. The shift transformation 
though does not depend on the number of $D3$ branes. So, by the same argument 
as before one expects to find the $SU(N)$ degrees of freedom
in the $adS$ side. However, as we argued in footnote \ref{nonab} it is 
rather unclear whether non-abelian singletons exist. So these degrees
of freedom should appear in a more subtle way.

In a parallel development Maldacena\cite{malda} (for earlier related work 
see \cite{kleb}) conjectured that the large 
$N$ limit of $\cn=4$ $SU(N)$ SYM is described by IIB supergravity on
$adS_5 \xx S^5$. This conjecture was further sharpened in 
\cite{GKP, witten2}, where a precise correspondence 
between the conformal field theory and the supergravity was proposed.
In particular,
it was shown that the chiral operators of $\cn=4$ $SU(N)$ SYM 
correspond to the Kaluza-Klein modes on $S^5$. This 
provides the resolution of the puzzle raised in the previous paragraph. 
Notice that the large $N$ limit is needed in our case in order to trust 
the spacetime picture. Similar remarks hold for the cases of $M2$ 
and $M5$. These developments have led to a series 
of papers \cite{toine}-\cite{rey}. 

Most of the recent papers were devoted to the study of $adS_4$,
$adS_5$ and $adS_7$. However, in these cases the corresponding 
superconformal theories to which the string theory (in the anti-de Sitter
background) is dual to are rather poorly understood. This prevents
detailed quantitative tests. In this paper we shall formulate
similar conjectures based on intersections of branes.
The corresponding dual superconformal field theories are two-dimensional
and depending on the particular intersection they may be chiral or not.
For the intersections we discuss the near-horizon limits correspond to
exact conformal field theories.
This opens the possibility to perform a detailed and
quantitative study of the conjectural relations between string 
theories on anti-de Sitter backgrounds and superconformal
field theories.

The new set of solutions of 11d and 10d supergravities 
are of the form $adS_k \xx S^l \xx S^m \xx E^n$,
with $k, l, m=2, 3$ and $n$ is such that the dimensions sum up to 
$10$ or $11$ depending on the solution. These solutions
are actually exact solutions of string theory, i.e. there 
is a conformal field theory (CFT) associated to them. For $S^3$ 
this is an $SO(3)$ WZW model and for $adS_3$ an $SL(2, R)$ WZW model.
The one associated with $S^2$ \cite{GPSmonopole}
follows from the fact that $S^3$ is a $U(1)$ bundle over $S^2$
(Hopf fibration). The CFT associated with $adS_2$ 
follows in a similar way through an appropriate quotient of the $SL(2, R)$ 
WZW model\cite{bertotti}.
The solutions that we construct 
have specific gauge fields and antisymmetric tensors needed in 
order to have a CFT. In some cases, however, one may $T$-dualize
the solution to obtain a new one with the same spacetime geometry 
but field strengths which are not the canonical ones from the CFT point 
of view. This form of the solution is sometimes more natural from 
the M-theory point of view. 

The developments in the last few years suggest that the basic solutions 
of $11d$ SUGRA, namely the membrane 
(M2), the fivebrane (M5), the wave (W) and 
the Kaluza-Klein monopole (KK) solution 
may form a ``basis'' in the space of all solutions of $11d$ SUGRA.
I.e. one may eventually obtain all solutions by 
appropriate (extremal or non-extremal) intersections of the basic ones,
dimensional reduction and use of dualities.
In support of this supposition we shall present  
intersecting brane configurations that correspond
to the new solutions. These intersections are obtained by 
appropriately combining solutions built according to 
the ``standard'' intersection rules\cite{PT, Tse1} with the solution of 
\cite{khuri, GKT}. In this way we obtain supersymmetric solutions 
that contain up to 6 different charges.

In all cases the intersection of branes interpolates between different 
stable vacua; Minkowski at infinity, 
$adS_k \xx S^l \xx S^m \xx E^n$ close to all branes 
and $adS_k \xx S^l \xx E^{n+m}$, $adS_k \xx S^m \xx E^{n+l}$,
close to some branes but far from the others. 
This is analogous to the case of M2, M5, D3, and NS5 studied 
in \cite{m5} and for ``standard'' intersections of branes 
studied in \cite{BPS1,KK}.

By explicitly computing the Killing spinors we find that 
in all cases the near-horizon solution exhibits doubling of supersymmetry.
Knowing how the Killing spinors transform under the isometry 
group allows us to determine the $adS$ supergroup that organizes 
the spectrum of the theory. The same supergroup can be interpreted as a 
conformal supergroup in one dimension lower. Based on these facts 
we argue that string theory on $adS_k \xx S^l \xx S^m \xx E^n$
backgrounds is dual to superconformal field theories in two dimensions.
For $k=3$ we find $(4,4)$ or $(4,0)$ SCFTs. For $k=2$ one would naively
expect one-dimensional $\cn=8$ superconformal theories, but we argue that
they are more properly viewed as Kaluza-Klein reductions of chiral SCFTs.
In all cases the SCFT contains the $\ca_\g$ algebra.
Our results are summarized in Table 3.

One may wonder whether one may use the shift transformation to reach 
these solutions.
In order to perform the shift transformation one needs to dualize 
each brane to a wave. For this to be possible each brane should 
be wrapped on a torus. However, the sphere parts of the solutions 
come from the worldvolume part of certain branes. So, if we want to end 
up with a solution that contains spheres we should not
wrap the brane on a torus. Spheres and also 
$adS$ spaces have non-abelian isometries. So, it may still
be possible to use non-abelian dualities to 
obtain an appropriate version of the shift transformation.

There is a simple rule (which we shall call the wave/monopole rule)
that leads to solutions with $adS_2$ and $S^2$'s once a solution with 
$adS_3$ and $S^3$'s is given. To get an 
$adS_2$ one adds a wave to $adS_3$, whereas 
to get an $S^2$ one adds a monopole to $S^3$.
One may anticipate these rules by looking at the corresponding CFTs.
To get an $S^2$ one views $S^3$ as a Hopf fibration over $S^2$.
The monopole solution precisely supplies the needed $U(1)$ 
monopole gauge field. Similarly, for $adS_2$ one needs a $U(1)$ gauge field
to support the $adS_2$ part. Putting a wave and reducing one
obtains a $D0$ brane that precisely provides the required gauge
field. The wave/monopole rule also allows for a determination of the 
killing spinors of all solutions once the killing spinors of the 
solutions containing only $adS_3$ and $S^3$'s are given.

We have organized the paper as follows. In the next section we present 
the wave/monopole rule. Using this rule we re-examine and organize
the solutions with near-horizon geometry $adS_k \xx S^l \xx E^m$ studied
in \cite{BPS1}.
Then we present the new solutions. These contain two 
sphere factors. This seems to be the maximum number since one can only
fit three 3-dimensional subspaces in $10$ dimensions (recall that 
the CFT of $adS_2$ and $S^2$ comes in terms of the CFT of $adS_3$ and $S^3$, 
respectively). In section 3, we analyze the supersymmetry enhancement 
of these configurations near the horizon. In particular, we explicitly
work out the case of $adS_3 \xx S^3 \xx S^3 \xx E^2$. We use these results 
in section 4 to argue that string theory on the background given by the 
new solutions is dual to certain superconformal theories.
In section 5, we present a study of 
the supersymmetry near the horizon of $D$-brane solutions in
arbitrary frames. We find  that the dual $Dp$-frame, i.e. the frame in 
which the curvature and the $(8{-}p)$-form field strength appear in the action
with the same power of the dilaton, represents a ``threshold''
frame for supersymmetry enhancement. In addition, in this frame 
the near-horizon geometry factorizes into the product
$adS_{p+2} \xx S^{8-p}$ for $p\neq 5$, whereas for $p=5$ it becomes
${\cal M}_7 \xx S^3$. In section 6, we use the new solutions 
to obtain several results about new vacua of gauged supergravities.
We also point out the possibility of a new gauged supergravity in
$5d$. Section 7 contains our conclusions. Finally, in the Appendix
we present a Kaluza-Klein ansatz inspired by the new solutions
and we explicitly compute the Killing spinors of the 
$adS_2 \xx S^2 \xx S^2 \xx E^5$ solution.

\section{Intersections with anti-de Sitter near-horizon limits}

In this section we describe special supersymmetric intersections of branes
which in their near-horizon limit have a factorized geometry involving
an anti-de Sitter spacetime and some other (compact) manifolds.
Most of them are found as solutions to $D=11$ supergravity
but in some cases we have to consider $D=10$ supergravity.
In \cite{BPS1} such solutions based on standard intersection
rules\footnote{By standard intersection rules we mean the rules
which give supersymmetric configurations in which the harmonic functions
depend only on overall transverse coordinates.}
were listed. Here we also use the non-standard intersection rule\footnote{
What we call a non-standard intersection of two branes
is often called an overlap, see \cite{Gaunt} for a motivation
of this nomenclature.}
which gives supersymmetric configurations in which the harmonic functions
depend on relative transverse coordinates only.
In $D=11$ the three standard intersection rules are
$(0|\,2\perp 2)$\footnote{
The notation $(q|\,p_1 \perp p_2)$ denotes a $p_1$-brane intersecting
with a $p_2$-brane over a $q$-brane.},
$(1|\,2\perp 5)$ and $(3|\,5\perp 5)$ \cite{PT, Tse1, GKT}, and the only
non-standard rule is $(1|\,5\perp 5)$ \cite{khuri, GKT}.
The intersection rules in ten dimensions can be derived from these
by dimensional reduction plus $T$ and $S$-duality.
For concreteness we give the standard and non-standard intersection
rules in the table below. References in which the intersection rules
are derived from the equations of motion are \cite{AEH} and \cite{ETT}
for standard and non-standard intersections, respectively.

\begin{table}[h]
\begin{center}
\vspace{.2cm}
\begin{tabular}{|c|c|c|}
\hline
          & standard            & non-standard       \\ \hline\hline
$D=11$    & $(0|M2\perp M2)$    &                    \\ \hline
          & $(1|M2\perp M5)$    &                    \\ \hline
          & $(3|M5\perp M5)$    & $(1|M5\perp M5)$   \\ \hline\hline
$D=10$    & $(\half(p+q-4)|Dp\perp Dq)$ & $(\half(p+q-8)|Dp\perp Dq)$\\ \hline
          & $(1|F1\perp NS5)$   &                    \\ \hline
          & $(3|NS5\perp NS5)$  & $(1|NS5\perp NS5)$ \\ \hline
          & $(0|F1\perp Dp)$    &                    \\ \hline
          & $(p-1|NS5\perp Dp)$ & $(p-3|NS5\perp Dp)$\\ \hline
\end{tabular}
\caption{{\it Standard and non-standard intersections in ten and eleven
 dimensions.}}
\end{center}
\end{table}

Since we use both standard and non-standard intersection rules, 
one has to specify which coordinates each harmonic function depends on.
In all cases this is clear by inspection of the intersection and it can be
further verified by looking at the field equation(s)
for the antisymmetric tensor field(s).

\subsection{Standard intersections}

There are three single $p$-branes with a near-horizon geometry
$adS_{p+2} \xx S^{D-p-2}$: the $M2$, $M5$ and $D3$ branes.
We now recall the solutions based on the standard intersections
and show that they are related by a simple rule which we will
call the wave/monopole rule.

\begin{table}[h]
\begin{center}
\begin{tabular}{|c|c|c|}
\hline
(a) & $M2\perp M5$ & $adS_3\times E^5\times S^3$\\ \hline
(b) & $M2\perp M2\perp M2$ & $adS_2\times E^6\times S^3$\\ \hline
(c) & $M5\perp M5\perp M5$ & $adS_3\times E^6\times S^2$\\ \hline
(d) & $M2\perp M2\perp M5\perp M5$ & $adS_2\times E^7 \times S^2$\\
\hline
\end{tabular}
\caption{{\it Standard intersection of M branes
 with anti-de Sitter near-horizon geometries.}} 
\end{center}
\end{table}

Table 2 shows the intersections and their near-horizon geometries
\cite{KK,BPS1}.
In \cite{BPS1} it was further demonstrated that the near-horizon limits of
these intersections exhibit supersymmetry doubling. Thus,
$M2\perp M5$ preserves $1/4$ and its near-horizon
limit $1/2$ of supersymmetry. The other three intersections
have supersymmetry enhancement from $1/8$ to $1/4$.

Roughly speaking, the wave/monopole rule asserts that starting
from the solution with near-horizon geometry containing $adS_3$
and $S^3$ factors (the $M2\perp M5$ intersection in table 2),
one can get all the others simply by adding a wave
and/or monopoles. A wave effectively replaces $adS_3$ by $adS_2$
and a monopole $S^3$ by $S^2$.
Let us see in some detail how this works for the wave.
The wave solution is given, in suitable coordinates
(light-cone coordinates), by the metric
\bea
ds^2 = K dx^2 + 2dxdt + dy_i dy_i\,,
\eea
where $K=K(y_i)$ is a harmonic function in the transverse space.
One can add such a wave in the direction of a common string
in the intersection (in these cases there is an $adS_3$ factor in
the near-horizon geometry). For $M2\perp M5$ plus a wave
the metric is\cite{Tse1}
\bea
ds^2&=&H_2^{-\frac{2}{3}}H_5^{-\frac{1}{3}}(K dx^2 + 2dxdt)
    +H_2^{-\frac{2}{3}}H_5^{\frac{2}{3}}(dx_2^2)
    +H_2^{\frac{1}{3}}H_5^{-\frac{1}{3}}(dx_3^2+\cdots +dx_6^2) \nonu
&&+H_2^{\frac{1}{3}}H_5^{\frac{2}{3}}(dr^2 + r^2 d\Omega_3^2)\,,
\eea
where $d\O_3$ is the line element on the unit three-sphere.
The harmonic functions are $H_{2,5}=1+{Q_{2,5}\over r^2}$ and
$K={Q_W\over r^2}$. Note that adding a constant to $K$ amounts
to a coordinate transformation $t\rightarrow t+\a x$.
In the near-horizon limit (or after applying the shift transformation)
the constants in the harmonic functions drop out and we obtain
\bea\label{adS3W}
ds^2 = dx^2 + d\rho^2 + 2 e^{2\rho}dxdt + ds^2_{E^5}+d\O_3^2\,,
\eea
where for convenience we have set $Q_2 = Q_5 = Q_W =1$, and we have
transformed to a new radial coordinate $\rho=\log{r}$.
Thus the metric splits up into a five-dimensional flat space,
a three-sphere and the first three terms which constitute three-dimensional
anti-de Sitter spacetime\footnote{This form of the $adS_3$ metric
was also used in \cite{CT} (and also appeared in \cite{Tse2})
with the same idea of reducing it to $adS_2$.}.
This is exactly the same near-horizon geometry as that of the
$M2\perp M5$ intersection without a wave. Adding the wave 
takes us to a different form of the $adS_3$ metric (provided that
no global identifications are made). To see this let us make the 
coordinate transformation\footnote{A similar coordinate transformation
was considered in \cite{hyun}.}
\be
x=x'-t', \quad t=\half(x'+t'), \quad e^{2\r} + 1 = r'^2.
\ee
This brings (\ref{adS3W}) to the form
\be \label{BTZ}
ds^2 = - {(r'^2 -1)^2 \over r'^2} dt'^2 + {r'^2 \over (r'^2 -1)^2} dr'^2
+r'^2 (dx'-{1 \over r'^2}dt')^2
\ee
This is the standard form of a massive extremal BTZ black hole
\cite{BTZ} with mass and angular momentum $M=J=2$
in a space of cosmological constant $\L=-1/l^2$ equal to $-1$.
These values simply reflect the fact that we have set all charges equal to 
$1$ in (\ref{adS3W}). Notice, however, that the coordinate $x'$ in
(\ref{BTZ}) is not periodic. It is well-known that the BTZ black hole
is locally isometric to $adS_3$, i.e. there is a coordinate transformation
that brings (\ref{BTZ}) to the anti-de Sitter metric. For non-extremal 
BTZ black holes this transformation is easy to write down.
In the extremal case the situation is more complicated and one needs to
consider an infinite number of patches that cover the black hole spacetime
(see \cite{BHTZ} section 3.2.4). Nevertheless this shows that
(\ref{BTZ}) with non-periodic coordinate $x'$ describes anti-de Sitter
spacetime. Therefore, so does (\ref{adS3W}). Note also that the wave does 
not contribute to the antisymmetric field strength.
Thus the near-horizon limit of $M2\perp M5 + W$ is the
same as that of $M2\perp M5$.
We conclude then also
that the supersymmetry of $M2\perp M5+W$ ($1/8$) is enhanced to $1/2$.
If we dimensionally reduce the metric (\ref{adS3W}) on $x$,
we get $adS_2$ (times $E^5 \xx S^3$) with a covariantly
constant two-form Kaluza-Klein field strength, and zero dilaton.
This type IIA solution is the near-horizon
limit of a $D0\perp F1\perp D4$ intersection. It 
preserves $1/4$ of supersymmetry since only half of the 
$adS_3$ Killing spinors survive the reduction.
This can be seen from the explicit form of the Killing spinors
as given in \cite{CH}. The surviving $adS_2$ Killing spinors
in the resulting coordinate frame are the ones given in \cite{LPT}.

The $D0\perp F1\perp D4$ configuration can be lifted
to a $D=11$ solution built only out of two and five-branes after first
$T$-dualizing along two relative transverse directions parallel to the
$D4$ brane. This yields $M2\perp M2\perp M2$ with near-horizon
geometry $adS_2 \xx E^6 \xx S^3$.
In the same way, $M5\perp M5\perp M5$ can be transformed
to $M2\perp M2\perp M5\perp M5$ by adding a wave.

To go from an $S^3$ to an $S^2$ one adds a monopole. This is based 
on the fact that $S^3$ can be written as a $U(1)$ bundle over
$S^2$ (Hopf fibration). The $U(1)$ field is precisely the monopole
field, explaining why we need a monopole in order to go from
a solution involving an $S^3$ to a solution involving an $S^2$.
Reduction or $T$-duality along the ``Hopf isometry'' of $S^3$
yields $S^2$ or $S^2 \xx S^1$, respectively.
Similar procedures were also described in the recent papers \cite{DLP,ITY}.
To illustrate the mechanism we consider again the $M2 \perp M5$ intersection
and add a monopole. This means that we replace the overall transverse 
space by a (single-center) euclidean Taub-NUT space (see \cite{EGH} p.363).
The solution is given by\cite{Tse1}
\bea
ds^2&=&H_2^{-\frac{2}{3}}H_5^{-\frac{1}{3}}(-dt^2 + dx_1^2)
    +H_2^{-\frac{2}{3}}H_5^{\frac{2}{3}}(dx_2^2)
    +H_2^{\frac{1}{3}}H_5^{-\frac{1}{3}}(dx_3^2+\cdots +dx_6^2) \nonu
&+&
H_2^{\frac{1}{3}}H_5^{\frac{2}{3}}
[H_M^{-1}(d\psi + Q_M \cos \theta d \f)^2 + H_M (dr^2 + r^2 d \O_2^2)]\,.
\eea
where $\psi$ is a periodic coordinate with period $4 \pi Q_M$,
$(\theta, \f)$ are coordinates on $S^2$ and
\be
H_M = 1 + {Q_M \over r}\ \,,\ \ \ H_{2,5}=1+{Q_{2,5}\over r}\,.
\ee
For small $r$ one may neglect the one in the various harmonic functions.
The result for the metric is (putting all charges but $Q_M$ to one)
\be
ds^2 = ds^2_{adS_3} + ds_{E^5}^2 + Q_M 
[(d\psi' + \cos \theta d \f)^2 + d\O_2^2]\,.
\ee
where $\psi'$ has period $4 \pi$.
This still represents $adS_3 \xx E^5 \xx S^3$, since the two terms
in the square brackets constitute $S^3$ in the coordinate system corresponding
to the Hopf fibration.
So again in the near-horizon limit nothing has changed by
adding a monopole.
Reducing along $\psi'$, one obtains the near-horizon limit of
$D2\perp NS5$ lying inside a $D6$ brane. Dualizing with
respect to two relative transverse coordinates of $D2$ and $D6$ one obtains 
a configuration of one $NS5$ brane and two $D4$ branes each of them 
intersecting the $NS5$ brane on a three brane. This configuration
can be uplifted to the standard $M5 \perp M5 \perp M5$ in $11d$.
The near-horizon geometry of the latter is $adS_3 \xx E^6 \xx S^2$.
In the same way the $M2 \perp M2 \perp M2$ solution may be 
transformed to $M2 \perp M2 \perp M5 \perp M5$ by adding a monopole.

The wave/monopole rule for the standard intersections
is summarized in the figure below.

\begin{figure}[h]
\begin{center}
\setlength{\unitlength}{1cm}
\begin{picture}(11,9)
\put(0,0){\makebox(4,1){$adS_2 \xx S^2$ (d)}}
\put(2,1){\line(0,1){2}}
\put(2,1){\line(2,1){4}}
\put(0,3){\makebox(4,1){$adS_2 \xx S^3$ (b)}}
\put(4,3){\makebox(4,1){$adS_3 \xx S^2$ (c)}}
\put(6,4){\line(0,1){2}}
\put(2,4){\line(2,1){4}}
\put(4,6){\makebox(4,1){$adS_3 \xx S^3$ (a)}}
\put(1.5,2){$M$}
\put(6.2,5){$M$}
\put(4,1.7){$W$}
\put(4,4.7){$W$}
\put(7.5,7.5){\makebox(4,1){$\#$ charges}}
\put(7.5,6){\makebox(4,1){$2$}}
\put(7.5,3){\makebox(4,1){$3$}}
\put(7.5,0){\makebox(4,1){$4$}}
\end{picture}
\setlength{\unitlength}{1pt}
\caption{{\it The wave/monopole rule for standard intersections. 
Starting from solution (a) one can obtain solutions (b), (c), (d)
by adding a wave and/or a monopole.
The lower-case letters in brackets refer to the four
standard intersections listed in table 2.}}
\end{center}
\end{figure}

\subsection{Non-standard intersections}

We now turn to intersections with near-horizon geometries
of the form $adS_k \xx S^l \xx S^m \xx E^n$ with $k,l,m$ equal to 2 or 3.
As we will see, all such configurations involve at least one pair of
branes intersecting in the non-standard way, and therefore
the harmonic functions will depend on relative transverse coordinates.
It was shown in \cite{CT} that the near-horizon limit
of the intersection of two ten-dimensional NS
five-branes over a line\footnote{
By near-horizon limit we mean here the asymptotic configuration
in the region near both five-branes.}
has geometry ${\cal M}_4 \xx S^3 \xx S^3$,
covariantly constant field strength and a linear dilaton.
Addition of a fundamental string along the
common direction of the five-branes yields a solution
with near-horizon geometry $adS_3 \xx S^3 \xx S^3 \xx E^1$
and constant dilaton, provided the
harmonic function of the string is the product of the
five-brane harmonic functions \cite{CT}.
In $D=11$ the analogous solution is the intersection
of two $M5$ branes over a line, plus a membrane:
$$
\begin{array}{lccccccccccc}
M5_1 & 1 &   & 3 & 4 & 5 & 6 &   &   &   &    & \\
M5_2 & 1 &   &   &   &   &   & 7 & 8 & 9 & 10 & \hspace{2cm} {\rm (A)} \\
M2   & 1 & 2 &   &   &   &   &   &   &   &    &
\end{array}
$$
where $H_{F}^{(1)}(x_7, x_8, x_9, x_{10})$, $H_{F}^{(2)}(x_3, x_4, x_5, x_6)$,
$H_{T}=H_{F}^{(1)} H_{F}^{(2)}$. (See (\ref{M552}) below for the
explicit solution in terms of the metric and antisymmetric tensor field.)
This solution preserves $1/4$ of supersymmetry.
The easiest way to check that one can take for
the membrane harmonic function the product of the five-brane
harmonic functions is to consider the field equation for
the antisymmetric tensor field, $\partial_M (\sqrt{-g} F^{MNPQ})=0$\footnote{
In general there is another term in this field equation coming
from the topological term in the $D=11$ supergravity action,
but this term vanishes in all configurations that will be
considered here.}. Substituting the metric associated to (A)
according to the harmonic function rule, one obtains (for
$N,P,Q$ in the membrane directions):
\bea\label{HT}
\left(H_F^{(1)}(x')\partial_x^2 + H_F^{(2)}(x)\partial_{x'}^2
\right)H_T(x,x')=0\,,
\eea
where $x$ denotes coordinates $x_3,\cdots x_6$ and $x'$
denotes $x_7,\cdots x_{10}$. Notice that (\ref{HT}) is the
Laplace equation in the curved transverse space produced by the
five-branes. Clearly, $H_T=H_F^{(1)} H_{F}^{(2)}$ is a solution.
What is essential though for obtaining the anti-de Sitter product geometry
is that the near-horizon limit of $H_T$ behaves as $1/r^2 r'^2$,
so one could take a more general solution of (\ref{HT}) with that property.
For example, $H_T = 1+{Q_T \over r^2 r'^2}$ would represent a
membrane intersecting on a string localized within both fivebranes.
More generally, one could take $H_T = 1+ {Q_T^{(1)} \over r^2} + {Q_T^{(2)}
 \over r'^2}+ {Q_T^{(3)} \over r^2 r'^2}$.
In the remaining of this article we shall, for simplicity,
use $H_T=H_F^{(1)} H_F^{(2)}$. 
The near-horizon limit of (A) has geometry $adS_3 \xx S^3 \xx S^3 \xx E^2$
and preserves $1/2$ of supersymmetry as we will show in the next section.
So there is supersymmetry doubling also in this case.

Now consider intersection (A) with harmonic functions
$H_F^{(1)}=1+{Q_1\over r'^2}$, $H_F^{(2)}=1+{Q_2\over r^2}$ and
$H_T=H_F^{(1)} H_F^{(2)}$ where $r^2=x_3^2+x_4^2+x_5^2+x_6^2$ and
$r'^2=x_7^2+x_8^2+x_9^2+x_{10}^2$.
We can consider various limits of this solution.
First, if we go very far from one of the fivebranes,
$r\rightarrow\infty$ ($r'\rightarrow\infty$),
we recover the standard intersection (a), i.e. $M2\perp M5$\footnote{
The same result is obtained by setting
$Q_2=0$ ($Q_1=0$).}. If we go near the horizon in the latter intersection,
$r'\rightarrow0$ ($r\rightarrow 0$), we find again the near-horizon
geometry $adS_3 \xx E^5 \xx S^3$. We describe in detail the interpolating 
structure of solution (A) in figure 2. 

\begin{figure}[h]
\begin{center}
\setlength{\unitlength}{1cm}
\begin{picture}(11,7)(0,0)
\put(0,0){\makebox(4,1){$(adS_3 \xx E^2 \xx S^3 \xx S^3)_{1/2}$}}
\put(3,1){\circle*{.2}}
\put(3,1){\vector(0,1){5.1}}
\put(3,1){\vector(1,0){4.9}}
\put(8,0.5){\makebox(3,1){$(adS_3 \xx E^5 \xx S^3)_{1/2}$}}
\put(1,5.8){\makebox(3,1){$(adS_3 \xx E^5 \xx S^3)_{1/2}$}}
\put(5,3){\makebox(1,1){$({\rm A})_{1/4}$}}
\put(6,3.5){\vector(1,0){1.9}}
\put(5.5,4){\vector(0,1){2}}
\put(5.9,3.9){\vector(1,1){2}}
\put(5,5.8){\makebox(1,1){$(\rm a)_{1/4}$}}
\put(8,3){\makebox(1,1){$(\rm a)_{1/4}$}}
\put(8,5.8){\makebox(1,1){$({\cal M}^{11})_1$}}
\put(6,.5){$r$}
\put(2.5,4){$r'$}
\end{picture}
\setlength{\unitlength}{1pt}
\caption{{\it Interpolating structure of solution (A).
Keeping one of the radial coordinates fixed while the other 
tends to infinity we recover solution (a), which itself interpolates
between $adS_3 \xx E^5 \xx S^3$ and Minkowski. In addition, 
in the limit $r$ and $r'$ go to zero, (A) approaches 
$adS_3 \xx E^2 \xx S^3 \xx S^3$. 
The horizontal and vertical axes correspond to a solution 
((A) with the $1$ removed from one of
the harmonic functions)
which interpolates between the supersymmetric vacua
with geometries $adS_3 \xx E^2 \xx S^3 \xx S^3$
and $adS_3 \xx E^5 \xx S^3$.
The subscripts denote the fractions of unbroken
supersymmetry.}}
\end{center}
\end{figure}
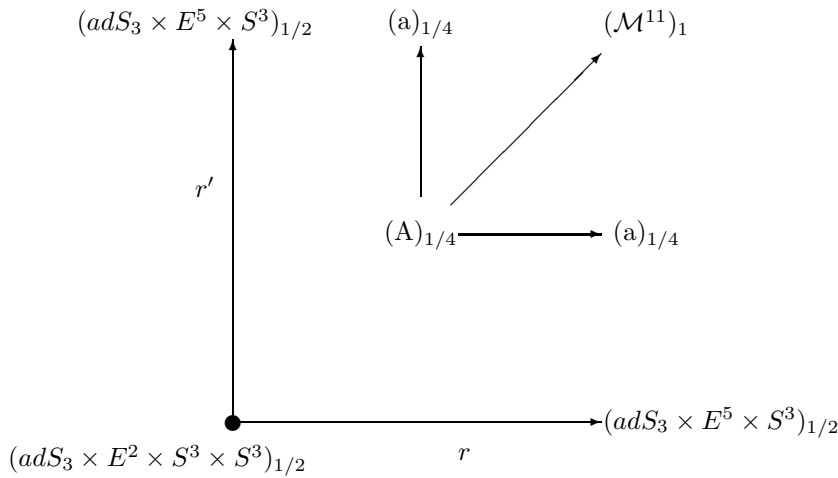

We can now use the wave/monopole rule to find other solutions
with near-horizon geometries where some or all of $adS_3 \xx S^3 \xx S^3$
are replaced by their two-dimensional versions.
The scheme is as follows:

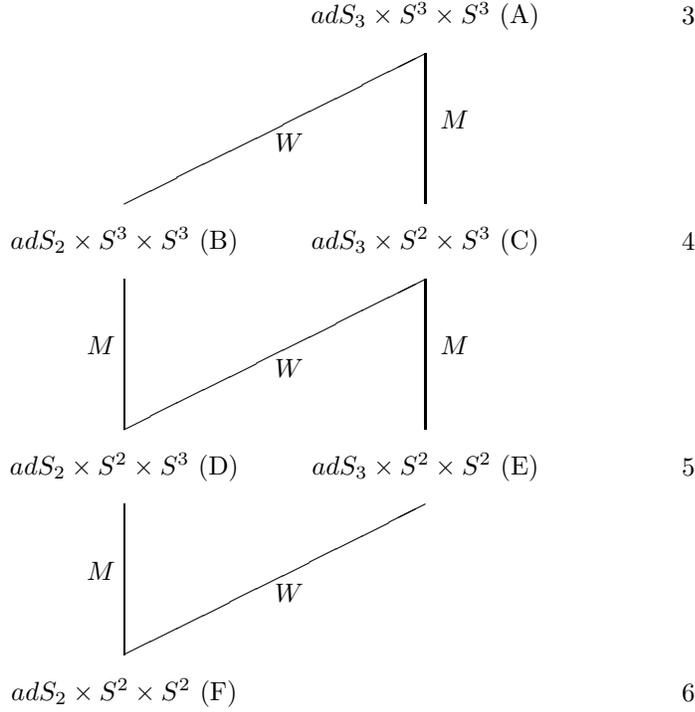
\begin{figure}[ht]
\begin{center}
\setlength{\unitlength}{1cm}
\begin{picture}(11,12)
\put(0,0){\makebox(4,1){$adS_2 \xx S^2 \xx S^2$ (F)}}
\put(2,1){\line(0,1){2}}
\put(2,1){\line(2,1){4}}
\put(0,3){\makebox(4,1){$adS_2 \xx S^2 \xx S^3$ (D)}}
\put(4,3){\makebox(4,1){$adS_3 \xx S^2 \xx S^2$ (E)}}
\put(6,4){\line(0,1){2}}
\put(2,4){\line(2,1){4}}
\put(4,6){\makebox(4,1){$adS_3 \xx S^2 \xx S^3$ (C)}}
\put(1.5,2){$M$}
\put(6.2,5){$M$}
\put(4,1.7){$W$}
\put(4,4.7){$W$}
\put(2,4){\line(0,1){2}}
\put(0,6){\makebox(4,1){$adS_2 \xx S^3 \xx S^3$ (B)}}
\put(6,7){\line(0,1){2}}
\put(2,7){\line(2,1){4}}
\put(4,9){\makebox(4,1){$adS_3 \xx S^3 \xx S^3$ (A)}}
\put(1.5,5){$M$}
\put(6.2,8){$M$}
\put(4,7.7){$W$}
\put(7.5,10.5){\makebox(4,1){$\#$ charges}}
\put(7.5,9){\makebox(4,1){$3$}}
\put(7.5,6){\makebox(4,1){$4$}}
\put(7.5,3){\makebox(4,1){$5$}}
\put(7.5,0){\makebox(4,1){$6$}}
\end{picture}
\setlength{\unitlength}{1pt}
\caption{{\it The wave/monopole rule for non-standard intersections. 
Starting from solution (A) one can obtain solutions (B), (C), (D), (E), (F)
by adding a wave and/or monopole(s).
The upper-case letters correspond to the solutions given in the text.}}
\end{center}
\end{figure}

They correspond to the following intersections.
Adding a wave to solution (A) and dimensionally reducing to type IIA
supergravity in the same way as described before for $M2\perp M5$,
one finds
$$
\begin{array}{lcccccccccc}
D0   &   &   &   &   &   &   &   &   &   & \\
D4_1 &   & 2 & 3 & 4 & 5 &   &   &   &   & \\
D4_2 &   &   &   &   &   & 6 & 7 & 8 & 9 & \hspace{2cm} (B) \\
F1   & 1 &   &   &   &   &   &   &   &   &
\end{array}
$$
where $H_f=H_0=H_4^{(1)} H_4^{(2)}$, $H_4^{(1)}(x_6, x_7, x_8, x_9)$ and 
$H_4^{(2)}(x_2, x_3, x_4, x_5)$\footnote{This schematic way of writing the
harmonic functions is only intended to give the dependence on the
coordinates, and one may take different charges for all
harmonic functions.}.
This solution is $1/8$ supersymmetric as follows from the set of
supersymmetry projection operators associated to the various branes
in the configuration.
Its near-horizon limit has geometry $adS_2 \xx S^3 \xx S^3 \xx E^2$
and preserves $1/4$ of supersymmetry. The dilaton is a constant
(depending on the charges) in this limit.
For illustrative purposes we shall for this case write down
explicitly the solution and its near-horizon limit.
According to the harmonic function rule we get
\bea
ds^2=H_f^{-1}(H_0 H_4^{(1)} H_4^{(2)})^{-\half}(-dt^2)
 +H_f^{-1}(H_0 H_4^{(1)} H_4^{(2)})^{\half}dx_1^2 \nonumber\\
+H_0^{\half}{H_4^{(1)}}^{-\half}{H_4^{(2)}}^{\half}(dx_2^2+\cdots +dx_5^2)
+H_0^{\half}{H_4^{(1)}}^{\half}{H_4^{(2)}}^{-\half}(dx_6^2+\cdots +dx_9^2)
\,,\nonu
H_{01I}=\partial_I H_f^{-1}\ \,,\ \ \ F_{0I}=\partial_I H_0^{-1}\,,\\
F_{1m'n'p'}=\e_{m'n'p'q'}\partial_{q'}H_4^{(1)}\ \,,\ \ \ 
F_{1mnp}=\e_{mnpq}\partial_q H_4^{(2)}\,,\nonu
e^{-2\phi}=H_f H_0^{-{3\over2}}(H_4^{(1)} H_4^{(2)})^{\half}\,,\nonumber
\eea
where $I$ runs over all $m\in\{2,3,4,5\}$ and $m'\in\{6,7,8,9\}$.
The harmonic functions are given in terms of $H_4^{(1)}=1+{Q_1\over r'^2}$
and $H_4^{(2)}=1+{Q_2\over r^2}$ with $r^2 = x_2^2 +x_3^2+x_4^2+x_5^2$
and $r'^2=x_6^2+x_7^2+x_8^2+x_9^2$.
In the near-horizon limit $r\rightarrow 0$ and $r'\rightarrow 0$ the
configuration becomes
\bea
ds^2&=&(Q_1 Q_2)^{-2}(rr')^4(-dt^2)+dx_1^2 +{Q_2\over r^2}(dr^2 +r^2 d\O_3^2)
 +{Q_1\over r'^2}(dr'^2 +r'^2 d\O_3'^2)\nonu
 &=&e^{-4A\rho}(-dt^2)+d\rho^2+d\l^2+dx_1^2+Q_2 d\O_3^2
 +Q_1 d\O_3'^2\,,
\eea
where in the second line the coordinate transformation
\be
\rho=-A^{-1}\log {rr' \over \sqrt{Q_1 Q_2}}\ \,,\ \ \  
\l=A^{-1}\left[\sqrt{{Q_2\over Q_1}} \log{r} 
- \sqrt{{Q_1\over Q_2}} \log{r'} \right], \quad
 A=\left({1\over Q_1}+{1\over Q_2}\right)^{\half}
\label{cot}
\ee
has been used. The resulting metric describes $adS_2 \xx E^2 \xx S^3 \xx S^3$.
The $(t, \r)$ part of the metric is the standard 
form of the $adS_2$ metric in horospherical coordinates. The negative
cosmological constant (the ``radius'' $R$ of $adS_3$) 
is equal to $\L=-R^2 =-Q_1Q_2/4(Q_1+ Q_2)$.
The dilaton vanishes in the limit, $e^{-2\phi}=1$. The field strengths
become
\be\label{fieldst}
H_{01\r}=F_{0\r}=-2A\e_{0\r}\ \,,\ \ \
F_{1\a'\b'\c'}=-2Q_1\e_{\a'\b'\c'}\ \,,\ \ \ 
F_{1\a\b\c}=-2Q_2\e_{\a\b\c}\,,
\ee
where $\e_{0\r}$, $\e_{\a,\b,\c}$ and $\e_{\a',\b',\c'}$ are
volume forms on $adS_2$, $S^3$ and $S^3$, respectively.
(We use the convention
that the epsilon symbols with tangent space indices are constants.)
The covariantly constant field strengths in (\ref{fieldst}) support $adS_2$ and
the two $S^3$'s, respectively. Each of the factors of the geometry
together with the corresponding field strength represents a CFT,
as discussed in the introduction.

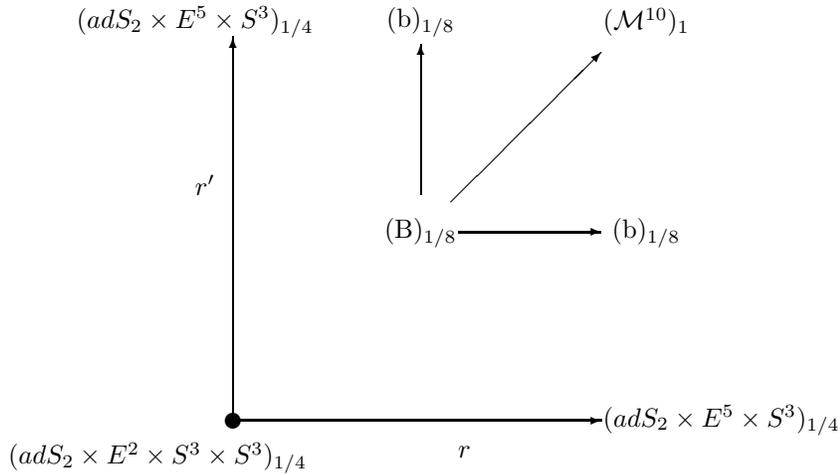
\begin{figure}[h]
\begin{center}
\setlength{\unitlength}{1cm}
\begin{picture}(11,7)(0,0)
\put(0,0){\makebox(4,1){$(adS_2 \xx E^2 \xx S^3 \xx S^3)_{1/4}$}}
\put(3,1){\circle*{.2}}
\put(3,1){\vector(0,1){5.1}}
\put(3,1){\vector(1,0){4.9}}
\put(8,0.5){\makebox(3,1){$(adS_2 \xx E^5 \xx S^3)_{1/4}$}}
\put(1,5.8){\makebox(3,1){$(adS_2 \xx E^5 \xx S^3)_{1/4}$}}
\put(5,3){\makebox(1,1){$({\rm B})_{1/8}$}}
\put(6,3.5){\vector(1,0){1.9}}
\put(5.5,4){\vector(0,1){2}}
\put(5.9,3.9){\vector(1,1){2}}
\put(5,5.8){\makebox(1,1){$(\rm b)_{1/8}$}}
\put(8,3){\makebox(1,1){$(\rm b)_{1/8}$}}
\put(8,5.8){\makebox(1,1){$({\cal M}^{10})_1$}}
\put(6,.5){$r$}
\put(2.5,4){$r'$}
\end{picture}
\setlength{\unitlength}{1pt}
\caption{{\it Interpolating structure of solution (B).}}
\end{center}
\end{figure}

The interpolation diagram for intersection (B) is shown in figure 4.
It is now understood that (b) corresponds to the standard
intersection $D0\perp D4\perp F1$ obtained from
$M2\perp M2\perp M2$ by a dimensional reduction
and two $T$-dualities. Since the reduction and $T$-dualities are
along relative transverse directions, one still has the $adS$
near-horizon geometry and the same number of unbroken supersymmetries.

The next solution is found by adding instead of a wave a monopole to solution
(A) and reducing to $D=10$, in the same way as illustrated before
for $M2\perp M5$. Notice, however, that one of the $M5$ branes lies
within the worldvolume directions of the additional monopole
and the other intersects on a string
with it. The latter is a non-standard intersection of a
monopole and a five-brane.
These and other intersection rules including waves and monopoles
are described in \cite{Ber}. The result in ten dimensions is
$$
\begin{array}{lcccccccccc}
D6  & 1 &   &   &   & 5 & 6 & 7 & 8 & 9 & \\
D4  & 1 & 2 & 3 & 4 &   &   &   &   &   & \\
NS5 & 1 &   &   &   & 5 & 6 & 7 & 8 &   & \hspace{2cm} (C) \\
D2  & 1 &   &   &   &   &   &   &   & 9 & 
\end{array}
$$
where $H_6=H_{s5}(x_2, x_3, x_4)$ and $H_2=H_{s5} H_4$,
$H_4(x_5, x_6, x_7, x_8)$.
This solution is $1/8$ supersymmetric.
The near-horizon limit has geometry $adS_3 \xx S^2 \xx S^3 \xx E^2$
and preserves $1/4$ of supersymmetry. The dilaton is a constant
in the limit. The interpolating structure of intersection (C)
is as follows ($r^2=x_2^2+x_3^2+x_4^2$ and $r'^2=x_5^2+x_6^2+x_7^2+x_8^2$):

\begin{figure}[h]
\begin{center}
\setlength{\unitlength}{1cm}
\begin{picture}(11,7)(0,0)
\put(0,0){\makebox(4,1){$(adS_3 \xx E^2 \xx S^2 \xx S^3)_{1/4}$}}
\put(3,1){\circle*{.2}}
\put(3,1){\vector(0,1){5.1}}
\put(3,1){\vector(1,0){4.9}}
\put(8,0.5){\makebox(3,1){$(adS_3 \xx E^4 \xx S^3)_{1/2}$}}
\put(1,5.8){\makebox(3,1){$(adS_3 \xx E^5 \xx S^2)_{1/4}$}}
\put(5,3){\makebox(1,1){$({\rm C})_{1/8}$}}
\put(6,3.5){\vector(1,0){1.9}}
\put(5.5,4){\vector(0,1){2}}
\put(5.9,3.9){\vector(1,1){2}}
\put(5,5.8){\makebox(1,1){$(\rm c)_{1/8}$}}
\put(8,3){\makebox(1,1){$(\rm a)_{1/4}$}}
\put(8,5.8){\makebox(1,1){$({\cal M}^{10})_1$}}
\put(6,.5){$r$}
\put(2.5,4){$r'$}
\end{picture}
\setlength{\unitlength}{1pt}
\caption{{\it Interpolating structure of solution (C).
Notice that in this case one obtains different standard intersections
((a) and (c)) depending on which radial coordinate is sent to infinity.}}
\end{center}
\end{figure}
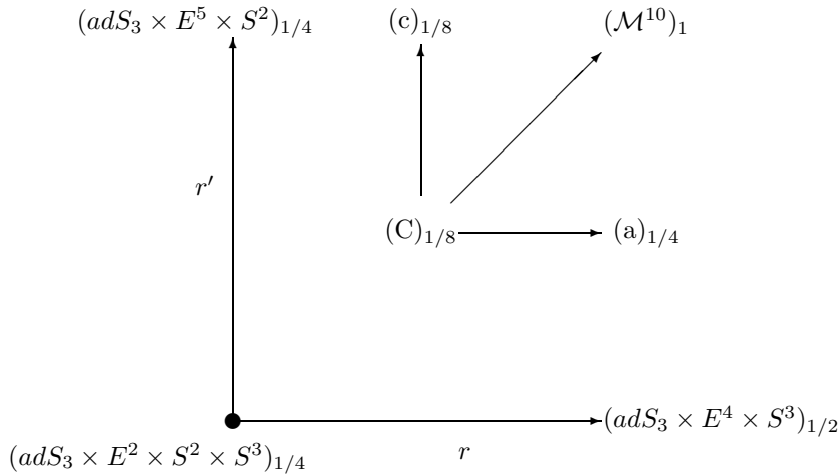

The type IIA solutions (B) and (C) cannot be lifted
to $D=11$
configurations with the same near-horizon geometry. In the standard
intersections this was always possible, but here
there is not enough freedom to dualize away the
Kaluza-Klein two-form gauge field without changing the horizon
geometry.

Whereas the four-charge configurations above are type IIA solutions,
we find that the five-charge configurations, if we want to write them
in terms of NSNS and/or RR branes only (without Kaluza-Klein monopoles
or wave), such that the near-horizon geometries are as indicated in
the scheme above,
can only be found as solutions of type IIB supergravity.
Adding a monopole to intersection (B) and applying $T$-duality along the
extra isometry direction of the monopole\footnote{
See also \cite{DLP} for a recent application of $T$-dualities along
isometries of odd-dimensional spheres.} we find:
$$
\begin{array}{lccccccccccc}
D1  & 1 &   &   &   &   &   &   &   &   & \\
D5  & 1 &   &   &   & 5 & 6 & 7 & 8 &   & \\
F1  &   &   &   &   &   &   &   &   & 9 & \hspace{2cm} (D) \\
NS5 &   &   &   &   & 5 & 6 & 7 & 8 & 9 & \\ 
D3  &   & 2 & 3 & 4 &   &   &   &   &   &
\end{array}
$$
where $H_f=H_1=H_3 H_5$ and $H_{s5}=H_5$, $H_3(x_5, x_6, x_7, x_8)$,
$H_5(x_2, x_3, x_4)$.
This solution can also be obtained by adding a wave to intersection (C)
and doing the $T$-duality along the wave direction $x$ as in (\ref{adS3W}).
It is $1/8$ supersymmetric, and the near-horizon limit has geometry
$adS_2 \xx S^2 \xx S^3 \xx E^3$ and preserves $1/4$ of supersymmetry.
Moreover, the dilaton is constant in this limit.
The interpolating structure of intersection (D)
is as follows ($r^2=x_2^2+x_3^2+x_4^2$ and $r'^2=x_5^2+x_6^2+x_7^2+x_8^2$):

\begin{figure}[h]
\begin{center}
\setlength{\unitlength}{1cm}
\begin{picture}(11,7)(0,0)
\put(0,0){\makebox(4,1){$(adS_2 \xx E^3 \xx S^2 \xx S^3)_{1/4}$}}
\put(3,1){\circle*{.2}}
\put(3,1){\vector(0,1){5.1}}
\put(3,1){\vector(1,0){4.9}}
\put(8,0.5){\makebox(3,1){$(adS_2 \xx E^5 \xx S^3)_{1/4}$}}
\put(1,5.8){\makebox(3,1){$(adS_2 \xx E^6 \xx S^2)_{1/4}$}}
\put(5,3){\makebox(1,1){$({\rm D})_{1/8}$}}
\put(6,3.5){\vector(1,0){1.9}}
\put(5.5,4){\vector(0,1){2}}
\put(5.9,3.9){\vector(1,1){2}}
\put(5,5.8){\makebox(1,1){$(\rm d)_{1/8}$}}
\put(8,3){\makebox(1,1){$(\rm b)_{1/8}$}}
\put(8,5.8){\makebox(1,1){$({\cal M}^{10})_1$}}
\put(6,.5){$r$}
\put(2.5,4){$r'$}
\end{picture}
\setlength{\unitlength}{1pt}
\caption{{\it Interpolating structure of solution (D).
In this case, similar to solution (C), one obtains different standard 
intersections ((b) and (d)) depending on which radial coordinate is sent 
to infinity.}}
\end{center}
\end{figure}
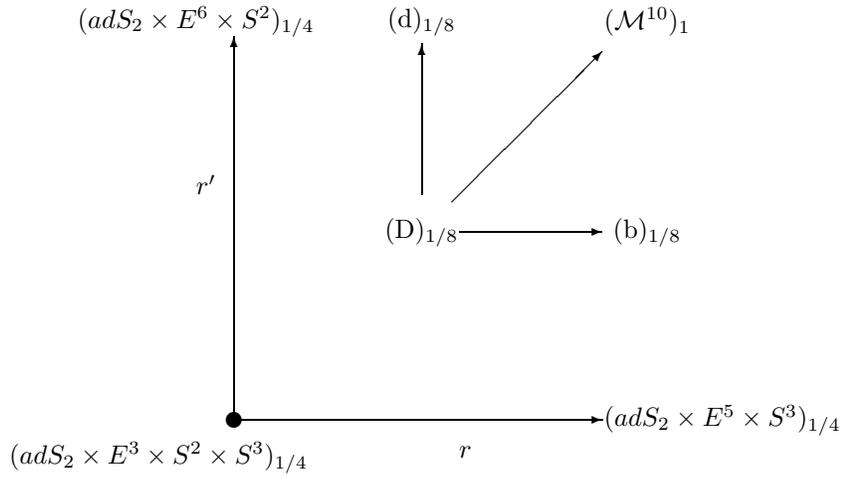

Adding a monopole to (C) and $T$-dualizing one gets the type IIB intersection
$$
\begin{array}{lccccccccccc}
D5_1  & 1 &   &   &   &   & 6 & 7 & 8 & 9 & \\
D5_2  & 1 & 2 & 3 & 4 & 5 &   &   &   &   & \\
NS5_1 & 1 &   &   &   & 5 & 6 & 7 & 8 &   & \hspace{2cm} (E) \\
NS5_2 & 1 & 2 & 3 & 4 &   &   &   &   & 9 & \\ 
D3    & 1 &   &   &   & 5 &   &   &   & 9 & 
\end{array}
$$
where $H_5^{(1)}=H_{s5}^{(1)}(x_2, x_3, x_4)$, 
$H_5^{(2)}=H_{s5}^{(2)}(x_6, x_7, x_8)$, 
$H_3=H_5^{(1)} H_5^{(2)}$.
It is $1/8$ supersymmetric.
Its near-horizon limit has geometry $adS_3 \xx S^2 \xx S^2 \xx E^3$
and preserves $1/4$ of supersymmetry. The dilaton is a constant in
this limit.
Note also that (D) and (E) are self-dual with respect to type IIB
$S$-duality.
The interpolating structure of intersection (E)
is given in figure 7 ($r^2=x_2^2+x_3^2+x_4^2$ and $r'^2=x_6^2+x_7^2+x_8^2$).

\begin{figure}[ht]
\begin{center}
\setlength{\unitlength}{1cm}
\begin{picture}(11,7)(0,0)
\put(0,0){\makebox(4,1){$(adS_3 \xx E^3 \xx S^2 \xx S^2)_{1/4}$}}
\put(3,1){\circle*{.2}}
\put(3,1){\vector(0,1){5.1}}
\put(3,1){\vector(1,0){4.9}}
\put(8,0.5){\makebox(3,1){$(adS_3 \xx E^5 \xx S^2)_{1/4}$}}
\put(1,5.8){\makebox(3,1){$(adS_3 \xx E^5 \xx S^2)_{1/4}$}}
\put(5,3){\makebox(1,1){$({\rm E})_{1/8}$}}
\put(6,3.5){\vector(1,0){1.9}}
\put(5.5,4){\vector(0,1){2}}
\put(5.9,3.9){\vector(1,1){2}}
\put(5,5.8){\makebox(1,1){$(\rm c)_{1/8}$}}
\put(8,3){\makebox(1,1){$(\rm c)_{1/8}$}}
\put(8,5.8){\makebox(1,1){$({\cal M}^{10})_1$}}
\put(6,.5){$r$}
\put(2.5,4){$r'$}
\end{picture}
\setlength{\unitlength}{1pt}
\caption{{\it Interpolating structure of solution (E).}}
\end{center}
\end{figure}

Finally, one can add a wave to (E) and $T$-dualize or add
a monopole to (D) and $T$-dualize to obtain the solution
with near-horizon geometry $adS_2 \xx S^2 \xx S^2 \xx E^4$.
Here there is again enough freedom to $T$-dualize in certain
relative transverse directions (corresponding to flat directions
in the near-horizon limit) such that we can uplift the solution to
an intersection of $M2$ and $M5$ branes with the same horizon
geometry. The resulting eleven-dimensional intersection
has already appeared in \cite{GGPT} and the configuration is as follows:
$$
\begin{array}{lccccccccccc}
M2_1 & 1 & 2 &   &   &   &   &   &   &   &    & \\
M2_2 &   &   & 3 & 4 &   &   &   &   &   &    & \\
M5_1 &   & 2 &   & 4 & 5 & 6 & 7 &   &   &    & \\
M5_2 & 1 &   & 3 &   & 5 & 6 & 7 &   &   &    & \hspace{2cm} (F) \\
M5_3 &   & 2 & 3 &   &   &   &   & 8 & 9 & 10 & \\
M5_4 & 1 &   &   & 4 &   &   &   & 8 & 9 & 10 & 
\end{array}
$$
where $H_{F}^{(1)}=H_{F}^{(2)}(x_8, x_9, x_{10})$, $H_{F}^{(3)}=
H_{F}^{(4)}(x_5, x_6, x_7)$,
$H_{T}^{(1)}=H_{T}^{(2)}=H_{F}^{(1)} H_{F}^{(3)}$.
This solution is $1/8$ supersymmetric as follows from the set of
supersymmetry projection operators associated to the various branes
in the configuration.
Its near-horizon limit has geometry $adS_2 \xx S^2 \xx S^2 \xx E^5$
and preserves $1/4$ of supersymmetry.
The interpolating structure of intersection (E)
is shown in figure 8 ($r^2=x_5^2+x_6^2+x_7^2$ and $r'^2=x_8^2+x_9^2+x_{10}^2$).

\begin{figure}[ht]
\begin{center}
\setlength{\unitlength}{1cm}
\begin{picture}(11,7)(0,0)
\put(0,0){\makebox(4,1){$(adS_2 \xx E^5 \xx S^2 \xx S^2)_{1/4}$}}
\put(3,1){\circle*{.2}}
\put(3,1){\vector(0,1){5.1}}
\put(3,1){\vector(1,0){4.9}}
\put(8,0.5){\makebox(3,1){$(adS_2 \xx E^7 \xx S^2)_{1/4}$}}
\put(1,5.8){\makebox(3,1){$(adS_2 \xx E^7 \xx S^2)_{1/4}$}}
\put(5,3){\makebox(1,1){$({\rm F})_{1/8}$}}
\put(6,3.5){\vector(1,0){1.9}}
\put(5.5,4){\vector(0,1){2}}
\put(5.9,3.9){\vector(1,1){2}}
\put(5,5.8){\makebox(1,1){$(\rm d)_{1/8}$}}
\put(8,3){\makebox(1,1){$(\rm d)_{1/8}$}}
\put(8,5.8){\makebox(1,1){$({\cal M}^{11})_1$}}
\put(6,.5){$r$}
\put(2.5,4){$r'$}
\end{picture}
\setlength{\unitlength}{1pt}
\caption{{\it Interpolating structure of solution (F).}}
\end{center}
\end{figure}

It is easy to check that all pairs of branes in the configurations (A)-(F)
satisfy either standard or non-standard intersection rules, see
table 1.
This concludes our presentation of the intersections with $adS$
near-horizon geometry. We have seen that adding a wave and/or monopoles
does not change the near-horizon limit of the intersection and
it is only after an additional dimensional reduction or $T$-duality
that one gets a different near-horizon geometry. For example, this
means that solution (A) with a wave and two monopoles added still
has supersymmetry enhancement to $1/2$ at the horizon.
A similar phenomenon also can be seen in the non-extremal generalizations
of certain intersections.
Let us illustrate this for the standard
intersection $M5\perp M5\perp M5$.
In the limit of keeping the non-extremality parameter $\m$ fixed
and taking large charges, the non-extremality function $f(r)=1-{\mu\over r}$
remains unchanged but the $1$'s in the harmonic functions of the
five-branes become negligible\footnote{
This limit can be reached by the shift transformation
without taking large charges \cite{KSKS}.
However, in that case the $adS$ part is really the non-extremal BTZ
black hole.}. The
geometry then becomes $adS_3 \xx E^6 \xx S^2$ (where $adS_3$ is
the non-extremal BTZ black hole without the identification, which is just
$adS_3$), and the field strengths are still covariantly constant w.r.t.
this metric. This is the same as the near-horizon limit of the extreme
version and therefore supersymmetry is (partially) restored in this limit.

\section{Supersymmetry enhancement}

A by now well-known phenomenon of certain solutions
with anti-de Sitter near-horizon geometry is supersymmetry
enhancement \cite{Gibb}.
For example, the $M2$, $M5$ and $D3$ branes
break one half of supersymmetry, whereas their near-horizon limits
are maximally supersymmetric vacua of $d{=}11$ supergravity.
These branes are therefore solitons which interpolate
between maximally supersymmetric vacua at the horizon and at infinity
\cite{m5}.
A few other examples of supersymmetry enhancement for static $p$-brane
solutions in different dimensions are known,
and in all these cases the near-horizon geometry contains a factor
$adS_{k}\times S^{m}$.
All solutions of table 2
exhibit supersymmetry enhancement at the horizon.
It turns out \cite{BPS1} that the condition for unbroken supersymmetry,
$\d\j_M =0$ where $\j_M$ is the eleven-dimensional gravitino,
in the background of the intersections in table 2, reduces
to the geometric Killing spinor equations on the anti-de Sitter,
flat and spherical factors of the geometry. In the case of
$M2\perp M5$ there is one additional projection, whereas
for the intersections with three and four charges there are two
projections needed.
As one projection reduces the supersymmetry by a factor one half
and since anti-de Sitter, flat and spherical geometries
all admit the maximal number of Killing spinors, one concludes
that the solutions in the right column of table 2 have double the
amount of supersymmetry as compared to their brane counterparts
in the left column.

Furthermore, for the configurations in table 2, a dimensional 
reduction over one or more of the relative transverse directions
(which correspond to the flat directions in the near-horizon limit)
will always give rise to lower dimensional solutions which also
exhibit supersymmetry enhancement at the horizon.
This is because all Killing spinors are independent of
these coordinates.
Further applications of $T$-duality in the
relative transverse directions or $S$-duality lead
to more solutions with supersymmetry enhancement.
The thus obtained class of solutions exhibiting
supersymmetry enhancement includes all previously known ones,
such as the four and five-dimensional extremal black holes
with nonzero entropy.
There are some other interesting features that all
these solutions have in common. They have regular (i.e. finite)
dilaton at the horizon (or no dilaton in eleven dimensions),
and in the shifted solutions the dilaton is a constant everywhere.
Besides, the antisymmetric field strengths become covariantly
constant in the shifted solutions, as in the Bertotti-Robinson solution.
Also, these solutions are non-singular \cite{GHT}.

The configurations (A) to (F), based on non-standard intersections,
turn out to have the same properties as those listed above for
the standard intersections. Near the horizon the dilaton has a fixed
finite value
and the field strengths become covariantly constant.
All of them have supersymmetry doubling as we will argue below.
We explicitly checked the Killing spinor equations for (A) and (F).
Solution (F) will
be discussed in detail in the appendix. Here we verify the
supersymmetry enhancement of (A).

The explicit solution belonging to configuration (A) is \cite{TG,Gaunt}
\bea
ds^2 = (H_T)^{\frac{1}{3}}(H_F^{(1)} H_F^{(2)})^{\frac{2}{3}}
 \big\{(H_T H_F^{(1)} H_F^{(2)})^{-1}(-dt^2 +dx_1^2)\nonu
 + (H_T)^{-1} dx_2^2 + (H_F^{(1)})^{-1}(dx_3^2 +\cdots +dx_6^2)
 + (H_F^{(2)})^{-1}(dx_7^2 +\cdots dx_{10}^2)\big\}\,,
\label{M552}\\
F_{012I}=-\partial_I (H_T)^{-1}\,,\ \ \ F_{2m'n'p'}=\e_{m'n'p'q'}
\partial_{q'} H_F^{(1)}\,,\ \ \ F_{2mnp}=\e_{mnpq}\partial_q H_F^{(2)}
\,,\nonumber
\eea
where $I$ runs over all $m\in\{3,4,5,6\}$ and $m'\in\{7,8,9,10\}$.
Going near the horizon, one can neglect the ones in the harmonic
functions, so $H_F^{(1)}={Q_1\over r'^2}$, $H_F^{(2)}={Q_2\over r^2}$
and $H_T ={Q_1 Q_2\over r^2 r'^2}$, where $r^2=x_3^2+\cdots +x_6^2$
and $r'^2=x_7^2+\cdots +x_{10}^2$.
The geometry then becomes
\bea
ds^2=(Q_1 Q_2)^{-1}r^2 r'^2 (-dt^2+dx_1^2)+dx_2^2+{Q_2\over r^2}dr^2
 + {Q_1\over r'^2}dr'^2
 +Q_2 d\Omega_3^2 + Q_1 d\Omega'^2_3\nonu
 =e^{-2A\rho}(-dt^2+dx_1^2)+d\rho^2 + 
 dx_2^2 +d\lambda^2
 +Q_2 d\Omega_3^2 + Q_1 d\Omega'^2_3\,,
\eea
where in the last line the
change of coordinates (\ref{cot}) has been performed,
as in \cite{CT}.
This is a metric for $adS_3 \xx E^2 \xx S^3 \xx S^3$.
The field strengths become
\bea
F_{\m\n\r 2}=2A\e_{\m\n\r}\,,\ \ \ 
F_{\a\b\c 2}=2Q_2\e_{\a\b\c}\,,\ \ \ 
F_{\a'\b'\c' 2}=2Q_1\e_{\a'\b'\c'}\,,
\label{fieldstr}
\eea
where $\mu\in\{0,1,\r\}$ is the $adS_3$ index and
$\a$ and $\a'$ are indices for the two $S^3$ factors, respectively.
The field strengths are covariantly constant.
Killing spinors are solutions of
\bea\label{KS}
\d\psi_M=D_M\e+{1\over288}(\G_M{}^{NPQR}-8\d_M{}^N\G^{PQR})F_{NPQR}\e=0\,.
\eea
The $\G$-matrices can be chosen
\bea
&&\G^\m = \c^\m\ot\c^3\ot\id\ot\id\ot\s_2\,,\nonu
&&\G^s=\id\ot\c^s\ot\id\ot\id\ot\s_2\,,\nonu
&&\G^\a=\id\ot\id\ot\c^\a\ot\id\ot\s_1\,,\\
&&\G^{\a'}=\id\ot\id\ot\id\ot\c^{\a'}\ot\s_3\,,\nonumber
\eea
where the index $s$ is used for the flat directions $x_2\,,\,\l$,
and $\c^3 =i\c^2\c^\l$.
The matrices $\c^\m\,,\,\c^s\,,\,\c^\a\,,\,\c^{\a'}$ are gamma-matrices
in $adS_3$, $E^2$, $S^3$ and $S^3$, respectively, and $\s_i$ are the
Pauli spin matrices.
Substituting in (\ref{KS}), one finds for the $E^2$ directions\footnote{
For notational convenience we have set $Q_1=Q_2=1$.}
\bea
\d\psi_s=\partial_s\e + {\d_{s\l}\over6}\left(-\sqrt{2}i\id\ot\id\ot
 \id\ot\id\ot\s_2 +\id\ot\c^3\ot\id\ot\id\ot(\s_1+\s_3)\right)\e\nonu
+ {\d_{s2}\over3}\left(\sqrt{2}\,\id\ot\c^3\ot\id\ot
 \id\ot\s_2+i\id\ot\id\ot\id\ot\id\ot(\s_1+\s_3)\right)\e\,.
\eea
Taking the spinors independent of $x_2,\l$, we must require
\bea\label{pr1}
\sqrt{2}\c^3\xi\ot\s_2\chi+i\xi\ot(\s_1+\s_3)\chi=0\,,
\eea
where we also wrote $\e=\eta\ot\xi\ot\r\ot\r'\ot\chi$.
This can also be written as
\bea
\label{proj}
{\cal P}(\xi\ot\chi)={1\over2}(1+\G)(\xi\ot\chi)=
 {1\over2}\left(1-{1\over\sqrt{2}}\c^3\ot(\s_1-\s_3)\right)(\xi\ot\chi)=0\,.
\eea
Since $\G$ is traceless and $\G^2=1$ this is a projection which
 breaks $1/2$ of supersymmetry.
For the $adS_3$ components of (\ref{KS}) we find
\bea
D_\m\e-{1\over6}\left(\c_\m\ot\c^\l\ot\id\ot\id\ot(\s_1+\s_3)
 +2\sqrt{2}\c_\m\ot\c^2\ot\id\ot\id\ot\s_2\right)\e=0\,.
\eea
Using (\ref{pr1}) one can rewrite this as
\bea\label{KE1}
D_\m\e-{1\over\sqrt{2}}(\c_\m\ot\c^2\ot\id\ot\id\ot\s_2)\e=0\,.
\eea
Similarly, the $\a$ and $\a'$ components of (\ref{KS}) give
rise, together with the projection (\ref{pr1}), to the following
equations:
\bea\label{KE2}
D_\a\e-{i\over2}(\id\ot\c^2\ot\c_\a\ot\id\ot\s_2)\e=0\,,\\
D_{\a'}\e-{i\over2}(\id\ot\c^2\ot\id\ot\c_{\a'}\ot\s_2)\e=0\,.
\label{KE3}
\eea
Now observe that $\c^2 \ot\s_2$ commutes with the projection
operator ${\cal P}$ in (\ref{proj}). This means that we can still
decompose our projected spinor space into $\pm$ eigenspinors
of $\c^2 \ot\s_2$. Thus for these eigenspinors equations (\ref{KE1})
to (\ref{KE3}) reduce to the geometric Killing spinor equations
on $adS_3$ and the three-spheres,
\bea
\label{KSE}
&& D_\m\eta \pm {1\over\sqrt{2}}\c_\m\eta =0\,,\\
&& D_\a\rho \pm {i\over2}\c_\a\rho =0\,,\\
&& D_{\a'}\rho' \pm {i\over2}\c_{\a'}\rho' =0\,.
\eea
Since anti-de Sitter spacetimes and spheres admit the maximal number
of Killing spinors, we conclude that the only projection is
(\ref{pr1}) and thus the near-horizon limit of solution (A)
is $1/2$ supersymmetric. Thus there is supersymmetry doubling
as compared to the intersection itself\footnote{
Our calculation also seems to show that if one changes
the orientation of the membrane (corresponding to
the other sign in the membrane's field strength contribution),
in which case the intersection breaks all supersymmetry, the
near-horizon limit is still $1/2$ supersymmetric.}.

Now for the other configurations (B) to (F), one has to reduce or
$T$-dualize along one or more directions, and
some Killing spinors will not survive this procedure.
In the transition from $adS_3$ to $adS_2$ one reduces
along the coordinate $x$ as in (\ref{adS3W}) and only
half of the Killing spinors survive this reduction. In fact, from
the analysis in \cite{CH} it is clear that this half corresponds
to the Killing spinors which solve the Killing spinor equation
(\ref{KSE}) with one definite choice of sign, say minus. Thus the
additional projection operator is effectively $\half(1-\c^2\ot\s_2)$.
This is the same for the spheres and therefore the supersymmetry
is reduced by one factor of $\half$ relative to (A) in all cases.
Thus, the near-horizon limits of (B) to (F) preserve $1/4$
of supersymmetry.

\section{Dual superconformal field theories}

We have argued in the introduction that the shift 
transformation\cite{BPS1} implies that the degrees of freedom 
of $D3$, $M2$ and $M5$ should
organize themselves into multiplets of $adS_5, adS_4$ and $adS_7$,
respectively. Supporting evidence has been provided by many recent
papers. The conjecture of \cite{malda} may now be reformulated 
as stating that in the large $N$ limit ($N$ is the number of coincident 
branes) only the worldvolume degrees of freedom of 
$D3$, $M2$ and $M5$, respectively, are relevant. Then M-theory
(or string theory) on the corresponding background is 
equivalent to a SCFT.
This is very similar to the case of Matrix theory\cite{matrix} where in the 
large $N$ limit strings and other branes decouple and
only the $D0$ degrees of freedom are relevant.
Actually this may be more than just a similarity. We have argued \cite{KSKS} 
that the actual symmetry of M-theory contains 
elements that connect compactifications on tori to compactifications
on spheres.  Thus, the Matrix formulation of M-theory on some tori 
may be equivalent to M-theory on some spheres. 

Most of the recent papers have studied the cases of $adS_5, adS_4$ and $adS_7$.
However, in these cases it is difficult to test the conjecture of 
\cite{malda} since our knowledge of the 
corresponding superconformal theories is rather limited. 
In contrast to the $d>2$ cases, two dimensional SCFTs are rather 
well understood. In this section we formulate conjectures similar to the ones 
in \cite{malda} but involving the solutions presented in this article.
This leads to $d=2$ SCFTs.
In addition, in our case the $\a'$ corrections are under control
since we have an exact conformal field theory for each
solution. This means that one does not need the large $N$ limit 
for the correspondence to hold.
Hence, one may perform a detailed quantitative analysis.

In the last section we computed the number of supersymmetries 
that remain unbroken in the near-horizon limit of our new solutions.
We shall now use these results to obtain the superalgebra 
that organizes the spectrum of the near-horizon theory.
The master solution from which all others follow is the one with
near-horizon geometry $adS_3 \xx S^3 \xx S^3 \xx E^2$. 
All others can be obtained from it by a combination of adding
a wave, monopoles and sending the radius of one of the spheres to infinity.
In addition, one can follow what happens to the supersymmetries 
after these operations. 

The isometry group of the near-horizon limit of solution (A) contains 
$SO(2,2) \xx SO(4) \xx SO(4)$ where the first factor is associated to 
$adS_3$ and the other two to the two spheres. The spectrum is thus
organized by a superalgebra which is a superextension of this
bosonic part. There are a number of different $d=3$ anti-de Sitter 
supergroups. The latter have been classified in \cite{d3ads}. 
Since $SO(2,2)(=SL(2, R) \xx SL(2, R))$ is not simple
the corresponding supergroup is in general a direct product $G_1 \xx G_2$,
where $G_1$ and $G_2$ are both superextensions of $SL(2,R)$.
Thus, giving only the bosonic algebra is not sufficient to fix 
the corresponding superalgebra. However, in most cases the superalgebra
is uniquely determined if one knows how the supercharges transform
under the bosonic algebra. 

Let us rewrite the bosonic algebra as 
$(SL(2, R)_1 \xx SU(2)_{1a} \xx SU(2)_{1b}) \xx 
(SL(2, R)_2 \xx SU(2)_{2a} \xx SU(2)_{2b})$.
Then from our calculation in the 
previous section we know that the supercharges transform as
$(2_1, 2_{1a}, 2_{1b}, 0_2, 0_{2a}, 0_{2b})$
and $(0_1, 0_{1a}, 0_{1b}, 2_2, 2_{2a}, 2_{2b})$.
It follows\cite{toinepr} that the corresponding superalgebra is 
$D^1(2, 1, \a) \xx D^1(2, 1, \b)$, $0<\a,\b\leq 1$. The bosonic 
subalgebra of $D^1(2,1,\a)$ is
$SL(2, R) \xx SU(2) \xx SU(2)$. The superalgebra $D^1(2,1,\a)$ has 
$8$ fermionic generators. In addition, $D^1(2,1,1)$
is isomorphic to $Osp(4|2)$. 
For $\a \neq 1$, $D^1(2,1,\a)$ differ from $Osp(4|2)$
in the way the $SU(2)$ generators enter in the right hand side 
of the anti-commutator of supersymmetries.

The group $D^1(2, 1, \a)$ is also a (finite dimensional) $d=2$ superconformal
group. The question is then whether it admits an infinite dimensional
extension. Indeed, this turns out to be the case. The corresponding
infinite dimensional SCFT, the $\ca_\g$ algebra ($\a=\g/(1-\g)$), 
was introduced in \cite{STV,moreag}. The algebra contains an affine 
subalgebra corresponding
to $SU(2) \xx SU(2) \xx U(1)$ and a set of four dimension-1/2 fields.
For unitary representations, the parameter $\a$ is related to the levels 
$k_+, k_-$ of the two affine $SU(2)$  algebras as
$\a=k_-/k_+$. The central charge is given by $c=6 k_- k_+/(k_-+k_+)$.
For $\a=1$ the algebra reduces to the standard large $\cn=4$ algebra.
For arbitrary $\a$ the algebra contains two $\cn=4$ subalgebras.
Each of them realizes the small $\cn=4$ algebra (which contains only 
one $SU(2)$ factor).

We therefore conjecture that M-theory on $adS_3 \xx S^3 \xx S^3 \xx T^2$
is equivalent to a $(4,4)$ $\ca_\g \xx \ca_{\g'}$ SCFT.
There is a canonical way to split the superalgebra into a left and a right
moving part. The full isometry group is generated by left and right 
translations on the group manifolds $SL(2,R)$ and $SU(2)$('s). So, 
we take for the left superalgebra the superextension of the isometry 
subalgebra generated by left translations and for the right one the 
superextension of the isometry subalgebra generated by right translations.
In abelian groups, however, left and right translations coincide.
So there are only two $U(1)$'s coming from the torus $T^2$,
but we also only need two $U(1)$'s; one for each $\ca_\g$.
Thus in this way we have completely geometrized the $N=4$ SCFT 
algebra. In addition, it seems natural that
left and right affine $SU(2)$'s have the same level $k_{\pm}$, 
since these $SU(2)$'s come from the same sphere in the $adS$ picture.
Then there is one parameter in the SCFT for each modulus of the solution (A). 
In particular, the radii of the two spheres correspond to the levels of the 
two affine $SU(2)$'s. In addition, deformations of the spheres
should correspond to marginal operators on the SCFT side. 
In particular, by ``squashing'' the sphere one may lose 
one $SU(2)$ and gain a $U(1)$ factor in the isometry group.
On the SCFT side the marginal operator would move us from the 
first entry in Table 3 to the second one, and so on.  

Each isometry gives rise to a gauge symmetry after dimensional reduction
(we keep all massive modes, so the issue of consistent truncation does 
not arise).
According to the analysis of \cite{witten2}, the 
bulk gauge fields should couple to the global currents of the SCFT 
in the boundary. In our case, we precisely have one global current for 
each gauge field. Thus, following \cite{witten2}, we propose
\be \label{part}
\left< \exp(\int_{M^2} j_a A_0^a) \right>_{CFT} = Z_{S} (A_0)
\ee 
where $j^a$ denotes collectively all currents, $A_0$ is the boundary 
value of the bulk gauge field $A$ (this can also be a graviton), and  
$M^2$ is either (a conformal completion of) two-dimensional 
Minkowski space or $S^2$ depending on whether we consider 
$adS_3$ or its Euclidean version. 
$Z_{S} (A_0)$ denotes the string theory partition function
in the background specified by the boundary values $A_0$.

Let us emphasize that the near-horizon geometries we consider
correspond to exact CFTs.
So one can go beyond the supergravity approximation, and
prove or disprove (\ref{part}) by explicitly computing both 
sides. We hope to report a detailed analysis in a future publication.

All other cases can now be obtained by using the monopole/wave rule.
We tabulate these results below.

\begin{table}[h]
\begin{center}
\begin{tabular}{|c|c|}
\hline
      $adS_3 \xx S^3 \xx S^3 \xx T^2$  & 
      $\ca_\g \xx \ca_{\g'}$
\\ \hline
      $adS_3 \xx S^2 \xx S^3 \xx T^2$  & 
      $\ca_\g \xx (Vir \xx \widehat{SU(2)} \xx \widehat{U(1)})$
\\ \hline
    $adS_3 \xx S^2 \xx S^2 \xx T^3$  & $\ca_\g \xx (Vir \xx \widehat{U(1)}^2)$
\\ \hline \hline
      $adS_2 \xx S^3 \xx S^3 \xx T^2$  & 
      $\ca_\g \xx \widehat{SU(2)} \xx\widehat{SU(2)}
       \xx \widehat{U(1)}$ \\ \hline
      $adS_2 \xx S^2 \xx S^3 \xx T^3$  & 
      $\ca_\g \xx \widehat{SU(2)}\xx \widehat{U(1)}^2$      \\ \hline
      $adS_2 \xx S^2 \xx S^2 \xx T^4$  & 
      $\ca_\g \xx \widehat{U(1)}^3$  \\ \hline
\end{tabular}
\caption{{\it Near-horizon geometries of solutions (A)-(F) and the 
corresponding SCFTs.}} 
\end{center}
\end{table}

Vir denotes the Virasoro algebra and the hat an affine algebra.
To obtain the result in the second entry one adds a monopole to the one in the 
first entry and then reduces over the nut direction. This has the
effect of projecting out the spinor $2_{2a}$. Thus, one is left with 
$(4,0)$ supersymmetry. In addition, adding a monopole and reducing yields
$S^2$ instead of $S^3$. Thus, the isometry group of the new solution
loses an $SU(2)$ factor and gains a $U(1)$ (that used to be the
fiber). Similar remarks apply to the third entry. Now, in addition to
$2_{2a}$, the $2_{2b}$ spinor is projected out, but this yields the same
overall projection. The theory still has $(4,0)$ supersymmetry.
The isometry group loses again one $SU(2)$ and gains a $U(1)$.

To obtain the last three entries one adds a wave to the solutions
involving $adS_3$. This has the effect of projecting out the 
spinor $2_2$. This projection eliminates half of the supersymmetries;
the same ones that the projections corresponding to
transitions from $S^3$ to $S^2$ eliminate.
Following the discussion
of the $adS_3 \xx S^3 \xx S^3 \xx T^2$ case one expects 
string theory on the background given
in the left column of the last three entries
of Table 3 to be equivalent to a
superconformal quantum mechanical model with global symmetry 
given in the right column.
However, the way we get $adS_2$ from $adS_3$ suggests
that the corresponding
quantum mechanical model is a Kaluza-Klein reduction of a
chiral CFT. Indeed, the entries in the right column
correspond to chiral $(4,0)$ SCFTs.
Similar conjectures may also be formulated for the 
standard intersections \cite{malda}. 

In some cases one may consistently truncate the massive modes.
This leads to gauged supergravities. In particular, one may consistently 
gauge only an $SU(2)$ part of the $SO(4)$ isometry group of $S^3$.
In this case one would need only one $SU(2)$ global current in the 
boundary. Indeed, the $\ca_\g$ algebra can be consistently
truncated to the small $N=4$ algebra that contains only one
$SU(2)$ factor. Thus we expect that gauged supergravities
arising from $S^3$ compactifications are related to the small
$N=4$ algebra.

Let us finish this section with some further remarks
about the AdS/CFT correspondence.
The right hand side of (\ref{part})
seems to depend only on the near-horizon geometry.
In general, one may have two {\it different} brane configurations
(not related by dualities)
whose near-horizon limits are related through dualities. 
For example, one can connect through
$T$-dualities all the entries in the left column in Table 3
but this cannot be done for the corresponding brane
intersections (they have different number of charges).
On the other hand, the SCFTs in the right column are different.
(Presumably the connection between the different SCFTs by
marginal operators is related to the connection 
of the near-horizon geometries by $T$-dualities.) 
So, one may need to refine the proposal (\ref{part})
so it distinguishes between different brane configurations
that have equivalent near-horizon limit. For instance,
if one only considers the supergravity limit all entries
in table 3 are different. So, the adS/CFT equivalence may only hold in
the large $N$ limit, even though one may reliably calculate
the right hand side of (\ref{part}) for any $N$.

\section{Branes in a dual frame}

We would like to discuss now the issue of supersymmetry enhancement in
$10$ dimensions, in arbitrary frames, for several brane
configurations. We will start with the case of one single $Dp$ brane
(with $p\leq 6$),
which for $p=1,2,3$ was already discussed in \cite{KK} in two specific
frames, the string frame and the Einstein frame. We will find
agreement with their results in these special cases.
The metric, dilaton and field strengths for a single Dp brane are, in
the string metric, given by \cite{HorStrom}
\bea
&&ds^2 = H_p(r)^{-1/2} (-dt^2 + dx_1^2 + \cdots + dx_p^2) +  H_p(r)^{1/2}
(dx_{p+1}^2  + \cdots + dx_9^2), \nonu
&&A_{01\ldots p} = H_p(r)^{-1} -1; \ \ e^{-2\f}=H_p^\frac{p-3}{2}  \nonu
&&H_p(r)= 1 + \frac{\cq}{r^{7-p}}; \ r^2=x^2_{p+1} + \cdots + x^2_9.
\label{Dpbrane}
\eea
The reason that
we study supersymmetry enhancement in arbitrary frames, is the
fact that, in the limit $r \to 0$, the exponent of the dilaton either
vanishes or diverges after 
the shift transformation (or equivalently when one approaches the
horizon).  Therefore rescalings of the metric by powers of
the string coupling will lead to different behaviour depending on the
rescaled metric. Only for $p=3$ the dilaton itself vanishes, the metric
factorizes as $adS_5 \xx S^5$, and we get supersymmetry
enhancement. In the following we will take $p \neq 3$.

In order to study the number of supersymmetries preserved by the
background configurations (\ref{Dpbrane}),
one has to find the number of solutions to
the vanishing of the supersymmetry variations of the dilatino and the
gravitino, which in the string frame are given by (see e.g. \cite{BBJ})
\bea
&&\d\l=\left( \g^\m \partial_\m\f \right)\varepsilon +
\frac{3-p}{4(p+2)!} e^\f F_{\m_1 \ldots \m_{p+2}} \g^{\m_1 \ldots
\m_{p+2}} \varepsilon^\prime_p \nonu
&&\d\j_\m=D_\m\varepsilon + \frac{(-1)^p}{8(p+2)!} e^\f F_{\m_1 \ldots
\m_{p+2}} \g^{\m_1 \ldots
\m_{p+2}} \g_\m \varepsilon^\prime_p.
\label{susyfermions}
\eea
Let us define
\be
g_{(\a)} = e^{-\a\f} g_s,
\label{ga}
\ee
where $g_s$ is the metric in the string frame (\ref{Dpbrane}). First
of all, it is easy to see that, in order to go to the `dual Dp brane
frame', in which both the curvature and the $(8-p)$-form field strength
$\frac{1}{(p+2)!}\e_{\m_1 \ldots \m_{10}} F^{\m_1 \ldots \m_{p+2}}$
appear in the action with the same power of the dilaton exponential,
one has to set $\a = \a_p = 
\frac{2}{7-p}$. Furthermore, only in this metric, after performing the
shift transformation, the geometry factorizes into the product
$adS_{p+2} \xx S^{8-p}$ for $p\neq 5$, whereas for $p=5$ it becomes
${\cal M}_7 \xx S^3$, similar to the case of the solitonic fivebrane
in the string metric \cite{m5}.

We will now show that this metric also corresponds to what we shall
call `threshold supersymmetry enhancement', that is, we get
supersymmetry enhancement in frames with $\a < \a_p$ ($\a > \a_p$) for
$p<3$ ($p>3$), and not when $\a \geq \a_p$ ($\a \leq \a_p$). As $\a =
\half$ corresponds to the metric in the Einstein frame, we see that
the results of \cite{KK} are contained in ours. 
To this end, we first work out the supersymmetry variation of the
dilatino, when we plug in the solution (\ref{Dpbrane}), written in an
arbitrary frame, and after performing the shift
transformation. We find
\be
\d\l \sim (3-p) r^{- \frac{p-3}{8} [(p-7)\a +2 ]} {\cal P} \varepsilon.
\label{deltal}
\ee
Here $\cal P$ is a projection operator that projects out half the
number of components of the spinor $\varepsilon$. From this equation
we can already conclude that indeed there will be no supersymmetry
enhancement (in the limit $r\to 0$) in frames with $\a \geq \a_p$ for
$p<3$, and $\a \leq \a_p$ for $p>3$. In the frames where we rescaled
with $\a < \a_p$ ($p<3$), or $\a > \a_p$ ($p>3$), we do find the
supersymmetry variation of $\l$ to vanish without use of the
projection operator in the limit $r\to 0$. In order to check that we
really get enhancement of supersymmetry in these frames in this limit,
we should also consider the supersymmetry variation of the gravitino. One
can show that also the $r$-dependence of this variation is
proportional to an identical factor, so that this is indeed the case.

We will not discuss here intersections of several $Dp$ branes in
general, but restrict ourselves to some interesting
observations. In particular, we shall discuss supersymmetry
enhancement for an arbitrary number of asymptotically flat and
orthogonally intersecting $D3$ branes. 

When there is only one $D3$ brane present, the geometry factorizes as
$adS_5 \xx S^5$ (after making the shift transformation), and we do get
supersymmetry enhancement. The case with two
$D3$ branes, intersecting along a string, can be related to the
intersection of an $M2$ brane and an $M5$ brane that we considered
before. These intersections are related by $T$-duality transformations
and a compactification along relative transverse directions only,
so the conclusions on
supersymmetry enhancement from the M-theory perspective directly
apply. So also in this case the geometry factorizes, as $adS_3 \xx S^3
\xx E^4$, and we do get supersymmetry enhancement. For three
$D3$ branes we do not get a factorization of the geometry, and also no
supersymmetry enhancement occurs (so such a configuration always
preserves only $1/8$ of the supersymmetry). One can show this either
by a direct computation, or by observing that this configuration can
only be obtained from an M-theory configuration which does have
supersymmetry enhancement (three $M5$ branes or three $M2$ branes) by
compactifying or $T$-dualizing along one of the $adS$ directions; also
in the latter case we need to compactify this direction, thereby
reducing the number of geometric Killing spinors by one half \cite{LPT},
so there
will be no supersymmetry enhancement. When we consider a configuration
of four $D3$ branes, the result can be related to the
$M2\bot M2 \bot M5 \bot M5$ configuration we studied before;
now this relation is again by making
$T$-duality transformations (and also compactifying) only along relative
transverse
directions. Therefore the geometry factorizes, as $adS_2 \xx S^2 \xx E^6$,
and we do get supersymmetry enhancement (to leave $1/4$ of the
supersymmetry unbroken), unless the orientations of the branes
are such that all supersymmetry is broken.
Finally, solutions
with more than four $D3$ branes, pairwise intersecting
over strings, are not asymptotically flat and there is no indication
of supersymmetry enhancement.

\section{Compactifications and (gauged) supergravity}

In this section we comment on the implications of our results
for supergravity theories in various dimensions.
First we consider dimensional reductions along the flat coordinates
in the near-horizon limits of the $adS$ solutions.
Each of the $D$-dimensional $adS_k \xx E^l \xx S^m \xx S^n$ solutions
($D=11$ or $D=10$) described in this paper can be dimensionally
reduced along
$p \leq l$ of the flat directions, giving a solution with
geometry $adS_k \xx E^{l-p} \xx S^m \xx S^n$ and the same
amount of supersymmetry (counting the number of spinor components) in
$(D-p)$-dimensional maximal Poincar\'e supergravity.
For $p=l$ and $n=0$ (standard intersections)
these can be interpreted as the near-horizon
limits of $4d$ and $5d$
extreme black holes and $5d$ and $6d$ extreme black strings.
For example, the $M2\perp M5$ intersection gives rise, upon reduction
along the five relative transverse directions, to
dyonic string solutions in six dimensions \cite{DFKR}
with near-horizon geometry $adS_3 \xx S^3$.
The solution $adS_3 \xx S^3$ is itself a
$1/2$ supersymmetric vacuum configuration of $\cn=4$ $6d$
supergravity. If the $M2$ and $M5$ charges are taken to be equal
one gets the self-dual string \cite{DL} which can also be embedded into
$6d$ $\cn=2$ chiral supergravity where it breaks only $1/2$ of
supersymmetry, and hence its near-horizon limit $adS_3 \xx S^3$
is a maximally supersymmetric vacuum.

As an example of reducing non-standard intersections along
flat directions, we note that the near-horizon
limit of solution (F) reduces to a $1/4$ supersymmetric
$adS_2 \xx S^2 \xx S^2$ solution in six dimensions.
However, in this case it is not clear whether
this is the near-horizon geometry of some $6d$ black hole
(or if there exists a solution that interpolates between this horizon
geometry and Minkowski spacetime at infinity),
since we can reduce the intersection (F) itself only along four
directions. The fifth flat direction in the horizon geometry is
not an isometry of the full intersection.
Reducing (F) along the four relative transverse directions ($1,2,3,4$),
one obtains a solution in $d=7$ with two radial coordinates and near-horizon
geometry $adS_2 \xx S^2 \xx S^2 \xx S^1$, and it would be
interesting to see whether this solution can be interpreted as
a black hole.

In addition we can deduce the existence of solutions
with a certain amount of supersymmetry after spontaneous
compactification on spheres. These compactifications are expected
to give rise to solutions of gauged supergravities.
Several of these results are well-known, such as
the spontaneous compactification of $11d$ supergravity on $S^7$,
giving gauged $\cn{=}8$ supergravity in $d{=}4$, and
the $adS_7\times S^4$ and $adS_5\times S^5$ solutions
of $d{=}11$ supergravity and type IIB supergravity. The anti-de Sitter
parts of these solutions are
maximally supersymmetric vacua of gauged maximal supergravities
in seven and five dimensions.
Below we list some further results\footnote{See also \cite{BPS1} and
\cite{CT}.} and some predictions
for supersymmetric vacua of gauged supergravities. We hope
to report on these issues in more detail in a future publication.
For the intersections discussed here we always find either $S^2$ or $S^3$
in the horizon geometry.
It seems there is not much known about gauged supergravities that
might result from Kaluza-Klein reductions over two-spheres,
probably due to the problem of making a consistent Kaluza-Klein ansatz
for compactifications on even dimensional spheres (see e.g. the discussion
in \cite{DNP}).
A consistent KK ansatz for $S^3$ compactification can be made by
introducing non-abelian vector fields for the $SU(2)$ group manifold.
The compactification of $10d$ type I supergravity on $S^3$ then
leads to the $SU(2)$ gauged $\cn=2$ $7d$ supergravity of \cite{SalSez}.
One can associate certain solutions of this gauged supergravity
to known spontaneous compactifications
of type I supergravity having an $S^3$ factor in the geometry.
So, for example, the ${\cal M}_7\times S^3$ solution
\cite{m5} of type I supergravity
corresponding to the fivebrane with shifted harmonic function (the
harmonic function without the $1$)
corresponds to the $1/2$ supersymmetric domain wall solution with linear
dilaton of gauged $\cn=2$ $d=7$ supergravity found in \cite{LPS}.
Of the intersections (a) through (F), only (a) and (A) can
be embedded into the type I theory\footnote{Of course, one can add a wave
and/or monopoles but this will not change the near-horizon
geometry, as we have argued. That will happen only after reduction
or $T$-duality.}.
 From solution (a) one deduces the existence of a $1/2$ supersymmetric
$adS_3 \xx E^4$ solution of $7d$ gauged supergravity, which is indeed known
\cite{HKL}. Furthermore, solution (A) predicts the existence of a
$1/2$ supersymmetric $adS_3 \xx S^3 \xx E^1$ solution.
Once we dimensionally reduce $\cn=2$ $7d$ gauged supergravity
to lower-dimensional gauged supergravity there will be many
more such vacuum solutions, since as we have seen this reduction
can be done along isometries of $adS_3$ or $S^3$ yielding
$adS_2$ or $S^2$, respectively, with a covariantly constant
field strength but without introducing a dilaton.
For example, in six-dimensional
gauged supergravity we predict the existence of the following
supersymmetric vacua: $adS_2 \xx E^4$, $adS_3 \xx E^3$,
$adS_2 \xx S^3 \xx E^1$, $adS_3 \xx S^2 \xx E^1$ and $adS_3 \xx S^3$.

For the maximally supersymmetric theories, a Kaluza-Klein reduction of
$11d$ supergravity on the $S^3$ group manifold yields $SU(2)$ gauged
$\cn=2$ $8d$ supergravity \cite{SalSez2}. Recalling the near-horizon
geometries in $D=11$ containing an $S^3$ factor, we predict the
existence of $adS_3 \xx E^5$, $adS_2 \xx E^6$ and $adS_3 \xx S^3 \xx
E^2$ vacua, preserving $1/2$, $1/4$ and $1/2$ of supersymmetry,
respectively. Dimensional reduction of this theory gives rise to
maximally supersymmetric gauged supergravities in lower dimensions.
Similar to the discussion in the previous paragraph, we can construct
vacuum solutions of these supergravities. For instance, in seven
dimensions we obtain the following supersymmetric vacua: $adS_3 \xx
E^4$, $adS_2 \xx E^5$, $adS_3 \xx S^3 \xx E^1$, $adS_3 \xx S^2 \xx
E^2$, $adS_2 \xx S^3 \xx E^2$ and $adS_2 \xx S^2 \xx E^3$ (after an
extra dimensional reduction and $T$-duality).

Finally we consider compactifications on $S^3 \xx S^3$.
An explicit reduction of $10d$ type I supergravity
on the group manifold $S^3 \xx S^3$ was done in \cite{Cham}.
The resulting $4d$ theory is $\cn=4$ $SU(2) \xx SU(2)$ gauged
supergravity \cite{FS}. As observed in \cite{CT},
solution (A), whose type I supergravity analogue
represents a string localized on the intersection of two
fivebranes, implies that there is an $adS_3 \xx E^1$ vacuum.
The half gauged version of $\cn=4$ gauged supergravity,
where one of the $SU(2)$ coupling constants is set to zero,
corresponds to a compactification on the group manifold $S^3\xx T^3$,
and is therefore just a dimensional reduction of
$SU(2)$ gauged $7d$ supergravity \cite{Cowd}.
This version should have in addition the supersymmetric
vacua $adS_2 \xx E^2$ and $adS_2 \xx S^2$.
It would be interesting to see how these brane inspired solutions
are related to several known solutions of $4d$ gauged supergravity
\cite{FG,Singh}. The geometries do match, but the number
of unbroken supersymmetries appear not to.
For example, the $adS_2 \xx S^2$ Freedman-Gibbons solution
breaks all supersymmetry and is also a solution in the fully gauged
$SU(2) \xx SU(2)$ model.
In general, one expects more anti-de Sitter vacua to exist, but
there may not be so many supersymmetric (and therefore stable) ones.
For example, one easily finds $adS_2 \xx S^2 \xx E^6$,
$adS_2 \xx S^2 \xx S^2 \xx E^4$ and $adS_2 \xx S^2 \xx S^2 \xx S^2 \xx E^2$
solutions with constant dilaton of type I supergravity, by taking
appropriate covariantly constant field strengths supporting the
different Einstein spaces. A preliminary analysis suggests that
these solutions break all supersymmetry. This could be related to the
fact that they do not seem to have a simple brane interpretation.

Finally, let us remark that
an $S^3 \xx S^3$ reduction of $11d$ supergravity is expected to
give rise to a maximally supersymmetric $SU(2)\xx SU(2)$ gauged $5d$
supergravity, which to the best of our knowledge has not
been constructed. This theory should also be obtainable after a
consistent reduction on $S^3$ of the maximal $\cn=2$, $d=8$ gauged
supergravity.

\section{Conclusions}

In this paper we constructed several intersections of branes with
the special property that the near-horizon geometry has the
form $adS_k \xx E^l \xx S^m \xx S^n$ where $k,m,n$ can take the
values $2,3$. All these configurations involve the non-standard
intersection rule in which the number of common worldvolume directions
is two less than in the corresponding standard intersection.
The two spheres in the geometry are associated with two sets of relative
transverse coordinates.
One can derive all these solutions from the $1/4$ supersymmetric $11d$
intersection with near-horizon geometry
$adS_3\xx E^2\xx S^3\xx S^3$ describing a membrane intersecting on
the common string of two fivebranes.
Adding a wave does not change the near-horizon limit but if one
reduces (or in $d=10$ $T$-dualizes) along an appropriate isometry
direction of $adS_3$ one
obtains an intersection with near-horizon geometry in which
$adS_3$ is replaced by $adS_2$.
In the same way adding a monopole (plus reduction or $T$-duality)
replaces one $S^3$ by an $S^2$ near the horizon. Further
solutions may be obtained from the ones that contain two monopoles 
by replacing the two-monopole part with a toric Hyperk\"ahler 
manifold\cite{GGPT}.

The intersections we obtained have an interesting structure
in the sense that they interpolate between three or four different
vacua. Besides flat spacetime at infinity and the near-horizon limit
which corresponds to the limit of small radii in both relative
transverse spaces, one can also let one radius become small
and the other large in which case the solution becomes a vacuum
solution whose geometry is the near-horizon geometry with the appropriate
sphere replaced by flat space.
In fact, letting one of the radii go to infinity reproduces one
of the four standard intersections with near-horizon geometry
$adS_k\xx E^m\xx S^n$.

The near-horizon limits of the intersections that we obtained
imply the existence of certain vacua of gauged supergravities,
as we described in section 6.
Also, they provide an intersecting brane interpretation of certain
known such vacua. It would be interesting to look for
brane interpretations of other known solutions too.

By an explicit computation of the Killing spinors
we have shown that all $adS$ intersections exhibit a doubling
of unbroken supersymmetry near the horizon.
This is of importance for the association of a superconformal
theory to the string theory on the $adS$ background.
By considering the transformation properties of the 
unbroken supersymmetries under the isometry group of the background 
geometry, we argued what the superconformal groups of the dual 
superconformal theories should be. In all cases the superconformal algebra 
contains the large $\cn=4$ algebra.
We argued that the maximally
symmetric case $adS_3 \xx S^3 \xx S^3 \xx T^2$ corresponds to 
$\ca_\g \xx \ca_{\g'}$. The spacetime geometry provides a geometric 
realization of this algebra.

In a similar way we proposed a dual superconformal theory 
for all other solutions. Actually these results also follow
from the monopole/wave rule once one starts from the 
case of $adS_3 \xx S^3 \xx S^3 \xx T^2$. In the case 
that the solutions contain $adS_2$ one would expect
the dual theory to be a $0+1$ dimensional superconformal theory.
However, we argued that these quantum mechanical models are just
reductions of chiral SCFTs.

Our results open the possibility to explicitly check the conjectural 
equivalence between string theory on an $adS$ background and
superconformal field theories. In the case at hand one may proceed as
follows. As a first step the conformal dimensions
of the ``massless'' representation of the $\ca_\g$ 
should match the masses of the Kaluza-Klein harmonics.
This would fix the relations between the moduli 
of our solutions and the parameters of the SCFT.
Once this is done one may proceed to explicitly 
compute both sides in (\ref{part}). This calculation seems tractable 
since on the left hand side we have the partition
function of some $\cn=4$ SCFT and on the right hand side 
there are also known CFTs associated with the $adS$ background.
We hope to report on this and related issues in the 
near future.

\section*{Acknowledgements}
We would like to thank Marco Billo, Andrea Pasquinucci and  
Toine Van Proeyen for discussions.

\appendix
\section{Kaluza-Klein ansatz and supersymmetry}

Here we discuss the near-horizon limit of solution (F).
We start from a Kaluza-Klein ansatz for an eleven-dimensional
supergravity solution with covariantly constant four-form field
strength and we wish to obtain the geometry $adS_2 \xx E^5 \xx
S^2 \xx S^2$. The ansatz for the field strength is as follows:
\bea
&&F_{\m\n 12}=c_1\e_{\m\n}\,,\ \ \ F_{\m\n 34}=c_1\e_{\m\n}\,,\nonu
\label{KKa}
&&F_{\a\b 13}=c_2\e_{\a\b}\,,\ \ \ F_{\a\b 24}=-c_2\e_{\a\b}\,,\\
&&F_{\a'\b' 14}=c_3\e_{\a'\b'}\,,\ \ \ F_{\a'\b' 23}=c_3\e_{\a'\b'}
\,,\nonumber
\eea
where $\m,\n$ are $adS_2$ indices, $1,2,3,4$ are flat directions
(note that the fifth flat direction does not appear in the ansatz),
and $\a,\b;\a',\b'$ are indices for the two $S^2$'s.
The opposite sign of $F_{\a\b 24}$ is chosen in order
to obtain a supersymmetric configuration as will be shown below\footnote{
If one includes a sign for all six field strengths the supersymmetry
calculation gives two relations for the signs in order that supersymmetry
is not completely broken. This is directly related to
the fact that the corresponding intersection is supersymmetric
not for any possible orientations of the branes.}.
Substituting (\ref{KKa}) in the Einstein equation of $D=11$ supergravity,
\bea
R_{MN}={1\over12}(F_{MPQR}F_N{}^{PQR}-{1\over12}g_{MN}F^2)\,,
\eea
we get for the flat components of the Ricci tensor
\bea
R_{mn}=2\d_{mn}(-c_1^2+c_2^2+c_3^2)\,,
\eea
where $m,n$ run over $1,2,3,4$ and we have assumed that $g_{mn}=\d_{mn}$.
Then we require that $R_{mn}=0$ and thus
\bea\label{cond}
c_1^2=c_2^2+c_3^2\,.
\eea
The other nonzero components of the Ricci tensor then become
\bea
&&R_{\m\n}=-c_1^2 g_{\m\n}\,,\nonu
&&R_{\a\b}=c_2^2 g_{\a\b}\,,\\
&&R_{\a'\b'}=c_3^2 g_{\a'\b'}\,.\nonumber
\eea
It is now clear that we can take $adS_2$ and two $S^2$ factors
to solve the above Einstein equations.
The field equation for $F$ is automatically satisfied because
$F$ is covariantly constant and in (\ref{KKa}) there is always
at least one common index in any pair of $F$'s so that the topological term
$F\wedge F$ does not contribute to the field equation (see also the
footnote referred to just above equation (\ref{HT})).

Note that the above ansatz shows that a four-form field strength
not only can support a four-dimensional Einstein space, or its
dual a seven dimensional one, as in \cite{FR,GoH}, but that by taking some
of the components along flat directions we can also obtain
products of lower-dimensional Einstein spaces.

Next we check the supersymmetry of this solution.
The factorization of $\G$-matrices can be takes as
\bea
&&\G^\m=\c^\m\ot\id\ot\id\ot\id\,,\nonu
&&\G^s=\c^3_{adS}\ot\c^s\ot\c^3_S\ot\c^3_{S'}\,,\nonu
&&\G^\a=\c^3_{adS}\ot\id\ot\c^\a\ot\id\,,\\
&&\G^{\a'}=\c^3_{adS}\ot\id\ot\c^3_S\ot\c^{\a'}\,,\nonumber
\eea
where $\c^3_{adS}={1\over2}\e_{\m\n}\c^{\m\n}$,
$\c^3_S={i\over2}\e_{\a\b}\c^{\a\b}$ and $\c^3_{S'}={i\over2}
 \e_{\a'\b'}\c^{\a'\b'}$ such that $(\c^3_{adS})^2 =(\c^3_S)^2
 =(\c^3_{S'})^2=\id$. The index $s$ is used for all five flat directions.
Substituting this and (\ref{KKa}) into
the supersymmetry variation of the gravitino (\ref{KS}),
we find for the flat components
\bea
\d\psi_s=\partial_s\e+{1\over12}{\big\{}
 c_1\id\ot(\c_s{}^{12}+\c_s{}^{34})\ot\c^3_S\ot\c^3_{S'}+
 ic_2\c^3_{adS}\ot(\c_s{}^{13}-\c_s{}^{24})\ot\id\ot\c^3_{S'}
 \nonumber\\
\label{flat}
 +ic_3\c^3_{adS}\ot(\c_s{}^{14}+\c_s{}^{23})\ot\c^3_S\ot\id
 -4c_1(\d_{[s}^1\d_{t]}^2+\d_{[s}^3\d_{t]}^4)\id\ot\c^t\ot\c^3_S\ot\c^3_{S'}\\
 -4ic_2(\d_{[s}^1\d_{t]}^3 -\d_{[s}^2\d_{t]}^4)\c^3_{adS}\ot\c^t
  \ot\id\ot\c^3_{S'}
 -4ic_3(\d_{[s}^1\d_{t]}^4 +\d_{[s}^2\d_{t]}^3)\c^3_{adS}\ot\c^t
  \ot\c^3_S\ot\id{\big\}}\e\,.\nonumber
\eea
Since we take $\partial_s\e=0$, this equation reduces for $s=5$,
the flat direction which does not appear in the ansatz (\ref{KKa}), to
\bea
{\big(}c_1\id\ot(\c^{12}+\c^{34})\ot\c^3_S\ot\c^3_{S'}
 +ic_2\c^3_{adS}\ot(\c^{13}-\c^{24})\ot\id\ot\c^3_{S'}\nonumber\\
 +ic_3\c^3_{adS}\ot(\c^{14}+\c^{23})\ot\c^3_S\ot\id{\big)}\e=0\,.
\label{s5}
\eea
Taking $s=1,2,3,4$ in (\ref{flat}) one gets four equations which
reduce, after some algebra, to only two independent conditions:
\bea
&&{\cal P}_1\xi={1\over2}(1+\G_1)\xi=(1+\c^{1234})\xi=0\,,\nonu
&&{\cal P}_2\e={1\over2}(1+\G_2)\e=
 (1+i{c_2\over c_1}\c^3_{adS}\ot\c^{23}\ot\c^3_S\ot\id
 +i{c_3\over c_1}\c^3_{adS}\ot\c^{24}\ot\id\ot\c^3_{S'})\e=0\,,
\eea
where $\xi$ is the four-component spinor factor associated with $E^5$.
Both $\G_1$ and $\G_2$ square to one (using (\ref{cond}))
and are traceless.
The two projectors moreover commute and therefore break $3/4$ of supersymmetry.
Equation (\ref{s5}) is fulfilled after these projections.
One can also show that ${\cal P}_2$ corresponds to the projection
operator found for configuration (A), equation (\ref{proj}), and
${\cal P}_1$ corresponds to the extra projection due to the
reduction along isometries of $adS_3$ or $S^3$.

The other components of the Killing spinor equation are
\bea
&&\d\psi_\m=D_\m\e+{1\over12}{\big\{}ic_2\c_\m\id\ot(\c^{13}-\c^{24})\ot
 \c^3_S\ot\id\nonu
&&\ \ +ic_3\c_\m\ot(\c^{14}+\c^{23})\ot\id\ot\c^3_{S'}
-2c_1\c_\m\c^3_{adS}\ot(\c^{12}+\c^{34})\ot\id\ot\id{\big\}}\e=0\,,\\
&&\d\psi_\a=D_\a\e+{1\over12}{\big\{}c_1\id\ot(\c^{12}+\c^{34})\ot\c_\a
 \ot\id\nonu
&&\ \ +ic_3\c^3_{adS}\ot(\c^{14}+\c^{23})\ot\c_\a\ot\c^3_{S'}
 -2ic_2\c^3_{adS}\ot(\c^{13}-\c^{24})\ot\c_\a\c^3_S\ot\id{\big\}}\e=0\,,\\
&&\d\psi_{\a'}=D_{\a'}\e+{1\over12}{\big\{}c_1\id\ot(\c^{12}+\c^{34})\ot\c^3_S
 \ot\c_{\a'}\nonu
&&\ \ +ic_2\c^3_{adS}\ot(\c^{13}-\c^{24})\ot\id\ot\c_{\a'}
 -2ic_3\c^3_{adS}\ot(\c^{14}+\c^{23})\ot\c^3_S\ot\c_{\a'}\c^3_{S'}
  {\big\}}\e=0\,.
\eea
One can now reduce these equations to the respective geometric Killing
spinor equations by using identity (\ref{s5}).
One finds:
\bea
&&D_\m\e-{1\over2}c_1(\c_\m\c^3_{adS}\ot\c^{12}\ot\id\ot\id)\e=0\,,\\
&&D_\a\e-{i\over2}c_2(\id\ot\c^{13}\ot\c_\a\c^3_S\ot\id)\e=0\,,\\
&&D_{\a'}\e-{i\over2}c_3(\id\ot\c^{14}\ot\id\ot\c_{\a'}\c^3_{S'})\e=0\,.
\eea
Now the operators inside the brackets in the above three equations
commute with both projection operators ${\cal P}_1$ and ${\cal P}_2$,
and therefore we can decompose these equations according to
their $\pm i$ eigenvalues with respect to $\c^{12}$, $\c^{13}$
and $\c^{14}$, to obtain the geometric Killing
spinor equations on $adS_2$, $S^2$ and $S^2$, respectively,
\bea
&&D_\m\eta\mp{i\over2}c_1 \c_\m\c^3_{adS}\eta=0\,,\\
&&D_\a\rho\pm{1\over2}c_2 \c_\a\c^3_S \rho=0\,,\\
&&D_{\a'}\rho'\pm{1\over2}c_3 \c_{\a'}\c^3_{S'}\rho'=0\,,
\eea
where we wrote $\e=\eta(x^\m)\ot\xi\ot\rho(z^\a)\ot\rho'(z^{\a'})$.
We conclude that the solution preserves $1/4$ of supersymmetry.
The corresponding brane intersection, configuration (F),
is $1/8$ supersymmetric.

\end{document}